\documentclass[prb,superscriptaddress,showpacs]{revtex4-1}
\usepackage{graphicx}
\usepackage{dcolumn}
\usepackage{bm}
\usepackage{amsmath}
\usepackage{amssymb}
\usepackage{ulem}
\usepackage{color}

\newcommand{\kap}{Cu$_3$Zn(OH)$_6$Cl$_2$}

\def\X{{\cal X}}
\def\M{{\cal M}}

\def\S3{{$\sqrt3\times\sqrt3$}}

\def\exp{{\rm exp}}

\def\HT{{\rm HT}}
\def\INK{{\rm in\;K}\,\,}

\def\PPA{{\rm PPA}}
\def\spin{{\rm spin}}

\def\phonon{{\rm phonon}}

\def\spinHT{{\rm spin,HT}}
\def\CVspin{{C_V^{\rm spin}}}

\def\min{{\rm min}}
\def\max{{\rm max}}
\def\CV{{C_V}}

\def\JJJ{$J_1$-$J_2$-$J_d\,$}

\begin{document}

\title{Exchange energies of  kapellasite from high-temperature series analysis of the kagome lattice $J_1$-$J_2$-$J_d\,$-Heisenberg model.}
\author{B. Bernu}
   \affiliation{LPTMC, UMR 7600 of CNRS, Universit\'e Pierre et Marie Curie, Paris VI, F-75252 Paris Cedex 05, France}
\author{C. Lhuillier}
   \affiliation{LPTMC, UMR 7600 of CNRS, Universit\'e Pierre et Marie Curie, Paris VI, F-75252 Paris Cedex 05, France}
\author{E.~Kermarrec}
\affiliation{Laboratoire de Physique des Solides, Universit\'e Paris Sud 11, UMR CNRS 8502, F-91405 Orsay, France}
\author{F.~Bert}
   \affiliation{Laboratoire de Physique des Solides, Universit\'e Paris Sud 11,
UMR CNRS 8502, F-91405 Orsay, France}
\author{P.~Mendels}
   \affiliation{Laboratoire de Physique des Solides, Universit\'e Paris Sud 11,
UMR CNRS 8502, F-91405 Orsay, France}
   \affiliation{Institut Universitaire de France, 103 Boulevard Saint-Michel,
F-75005 Paris, France}
\author{R. H. Colman}
   \affiliation{University College London, Department of Chemistry, 20 Gordon Street, London, WC1H 0AJ, United Kingdom}
\author{A. S. Wills}
   \affiliation{University College London, Department of Chemistry, 20 Gordon Street, London, WC1H 0AJ, United Kingdom}
\date{\today}
\begin{abstract}
We present a method to build magnetic models for insulators based on high-temperature expansions by fitting both the magnetic susceptibility and the low temperature specific heat data.
It is applied to the frustrated magnet kapellasite (Cu$_3$Zn(OH)$_6$Cl$_2$) with the  $J_1$-$J_2$-$J_d\,$-Heisenberg model on the kagome lattice. 
Experimental data are reproduced with a set of \textit{competing} exchange energies closed to $J_1 = -12$\,K, $J_2= -4$\,K and $J_d=15.6$\,K, where $J_d$ is the third neighbor exchange energy across the hexagon.
Strong constrains between these exchange energies are established. 
These values confirm the results of B. F{\aa}k et al. (Phys. Rev. Lett., {\bf 109}, 037208 (2012)) regarding the location of kapellasite in the {\it cuboc2} phase of the Heisenberg model.
The quality and limits of this modeling are discussed.
\end{abstract}
\maketitle

\section{Introduction}
There are different routes for building magnetic models for insulators. The simplest and most reliable one is the modeling of inelastic modes (spin waves) as detected by neutron scattering, if any. In the case of a spin liquid, the inelastic spectrum is a continuum and may have very few distinct features when it is gapless. On the other hand, \textit{ab initio} calculations are notoriously difficult and strongly depend on the nature of the approximations. The only tool left is a modeling through fits of thermodynamic quantities to high temperature (HT) series. 
It is well known that the extraction of the Curie-Weiss temperature from susceptibility data  is quite delicate and requires a large range of high-temperature experimental data. In the case of frustrated magnets, this is insufficient to provide some insight in the low-temperature physics.\cite{FRUSTRES,FRUSTRES-2} In fact, as we will show in this paper, the fit of the susceptibility alone, even  in a large range of temperatures, does not settle the model and should be complemented by a fit of the magnetic specific heat.
This paper aims at unveiling the different difficulties that can be encountered in this process and can provide, with a given complex example, the case of kapellasite, a general method to tackle this problem. 

Kapellasite\cite{kapella1,kapella2} is a  polymorph of herbertsmithite and shares its chemical formula \kap. 
As for herbertsmithite, kapellasite fails to develop any long-range magnetic order down to 20 mK, displays a continuum of inelastic excitations, and is, thus, an interesting spin-liquid candidate.\cite{CUBOC2PRL} But contrary to herbertsmithite,\cite{helton07,Bert2007,MIS2007,rigol2007} the high-temperature susceptibility of this recently discovered metastable compound points to a ferromagnetic Curie-Weiss  field of about  10 K, whereas, the low-temperature behavior does not show  dominant ferromagnetic correlations down to the lowest temperature:  This observation is a characteristic of competing interactions. This compound is, thus, a delicate benchmark for any modeling, but it is also a very precious one as we know, from neutron-scattering data, it has very well defined and specific low-temperature short-range spin-spin correlations. Therefore, the results of the high temperature modeling can be immediately questioned through the low-temperature neutron data.\cite{CUBOC2PRL}

\begin{figure}
\includegraphics[width=0.25\columnwidth]{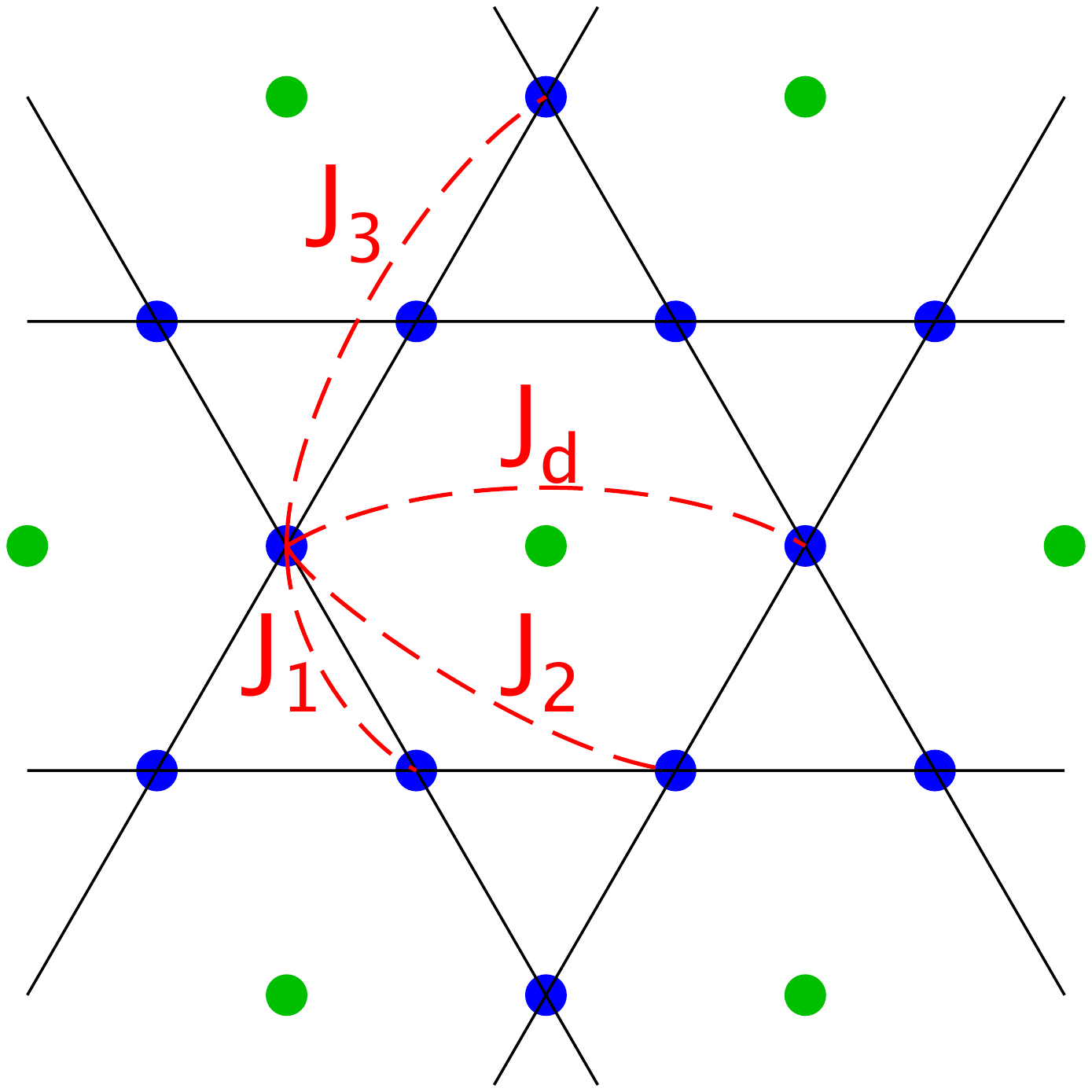}
\caption{(Color online) Kagome plane of kapellasite
with Cu$^{2+}$ $S\!=\!1/2$ spins (blue),
a non magnetic Zn$^{2+}$ ion (green),
and exchange interactions (red).
}
\label{FigLattice}
\end{figure}

While kapellasite has the same chemical formula as herbertsmithite, the two are not isostructural.  In kapellasite, the coupling between the kagome planes  occurs only via very weak O-H-Cl hydrogen bonds.\cite{kapella2} 
Kapellasite is, therefore, remarkably two dimensional.
A first theoretical description of kapellasite, which is deep in the Mott phase, is the Heisenberg Hamiltonian on the perfect kagome lattice,
\begin{equation}
	\mathcal{H} = \sum_{{\langle i,j\rangle}_\alpha} J_\alpha \, {\mathbf S}_i \cdot {\mathbf S}_j ,
\label{EqHberg}
\end{equation}
where the exchange integrals $J_\alpha$  are defined in Fig. \ref{FigLattice}.
Due to the  geometry of the exchange paths, $J_3$ and $J_d$ are different and
$J_d$ is expected to be larger than $J_3$ by an order of magnitude.\cite{Janson}
We will, thus, limit our analysis to the pure \JJJ model. 
We further neglect both the effects of disorder and of an eventual Dzyaloshinskii-Moriya (DM) interaction. 

The spin-1/2 HT series of magnetic susceptibility $\X$ and specific heat $C_V$  with the \JJJ parameters have been computed up to order  9 and are given in the Supplemental Material.\cite{SuppMat}
 
The paper is organized as follows.
In  Sec. II, the magnetic susceptibility $\X$ is fitted  to experimental data providing strong constraints on the coupling constants.
In Sec. III, we show how to use the low-temperature $C_V$  data to further refine these constraints.
Sections II and III are organized similarly. 
A quality factor is defined to measure the quality of the fits, whereas technicalities are reported in the appendices, and we finish with the physical conclusions to be kept in mind when considering the  properties of the model.
In the Conclusion, we discuss the consequences of neglecting, at this stage, the chemical disorder in the kagome plane  and Dzyaloshinskii-Moriya interactions.

\section{Describing $\X(T)$}
\label{SEC-chi}
The dc susceptibility was measured in a commercial Quantum Design MPMS-5S super conduction quantum-interference devoice (SQUID) magnetometer. It does not diverge at low temperature and coincides with the NMR local probe data indicating that the measured macroscopic susceptibility is intrinsic.
The experimental data are given as a list of points $\{T_k,\X_k^\exp\}$.
Fitting the data to a Curie-Weiss law $\X(T)\simeq C/(T-T_0)$ leads to $C=0.429(2)\,{\rm cm}^3\,{\rm K/mol}$ and $T_0=9.5\pm1$\,K, where $T_0$ is interpreted as the Curie-Weiss temperature.
In the range of temperatures of interest, $\X^{} T/C\simeq1$, thus, suitable for fitting.

We define the HT-series expansion of the magnetic susceptibility $\X^\HT$,
\begin{eqnarray}
	\frac{\X^{\HT}(T)T}{C}=1+\sum_{i=1}^n P_i(J_1,J_2,J_d) \beta^i,
\end{eqnarray}
where $\beta=1/T$ and  $P_i$ is a homogeneous polynomial of order $i$ and $n$ is the highest order at which the series is known.
The Curie-Weiss temperature is defined as $\theta=P_1(J_1,J_2,J_d)$ and, for the kagome lattice, $\theta=- J_1-J_2-J_d/2$.
These polynomials are given in the Supplemental Material\cite{SuppMat} up to order $n=17$,\cite{KAG-HT} 10, 11,  and 9 for the $J_1$ (M100), $J_1$-$J_2$ (M120), $J_1$-$J_d$ (M10d) and $J_1$-$J_2$-$J_d$ (M12d) models, respectively.

In order to account for the uncertainties in the number of spins and the temperature independent Van Vleck and diamagnetic susceptibilities, we introduce two parameters $A$  (close to 1) and $B$ and define a least mean square error as
\begin{eqnarray}
	\label{EQ-defZchi}
	Z_\X=\frac1{\epsilon^2N_T}\sum_{k=1}^{N_T}\left[A\frac{\X^\HT(T_k)T_k }{C} +BT_k -\frac{\X_k^\exp T_k }{C}\right]^2,\quad
\end{eqnarray}
where $N_T$ is the number of experimental measurements, $T_k\ge T_\min$ and $\epsilon=0.0015$ is on the order of the experimental uncertainties on $\X T/C$.
One could then minimize $Z_\X$ with respect to the parameters $\{J_1,J_2,J_d,A,B\}$.

Using Pad\'e approximants of the truncated series allows extension of  the fits to significantly lower temperatures and the definition of Eq.\ref{EQ-defZchi} is extended to $Z_{\X,\PPA}$ by replacing the HT polynomial by the various {\sl physical Pad\'e approximants} (PPAs) (see the definition in Appendix \ref{APP-PPA}).

In the present approach, the best set of parameters is searched among that having the largest number of Pad\'e approximants providing a {\it ``good fit"} of the experimental data. We, thus, define a measure $Q_\X$ of the fit quality as
\begin{eqnarray}
\label{EQ-Q-def}
	Q_\X&=&\sum_{\{\PPA\}} \M\left(Z_{\X,\PPA}\right),
\end{eqnarray}
where the sum runs over the PPAs and $\M(x)$ is a measure function chosen to be close to 1 for $x<1$ and to vanish rapidly for $x>1$ to discard bad PPAs. We use
\begin{eqnarray}
\label{EQ-def-M}
	\M(x)=\frac1{1+x^8}.
\end{eqnarray}
A ``good'' (respectively, ``bad'') PPA contributes 1 (respectively, 0) to $Q_\X$, thus, higher is the $Q_\X$, better is the fit.

The choice of $T_\min$: If  $T_\min$ is too high ($T_\min>25$\,K), almost all PPAs coincide with the HT polynomial,
and the experimental data do not strongly constrain the parameters of the model.
As $T_\min$ decreases, the constraint becomes stronger, but the PPAs start to deviate from each other, and the quality of the approximation becomes questionable.
This is seen in the function $Q_\X(T_\min)$, which decreases sharply around some $T_s$:  In the following $T_\min$ is chosen just above $T_s$.

We look for the set $\{J_1,J_2,J_d,A,B\}$ maximizing $Q_\X$. 
The evaluation of the linear parameters $A$ and $B$ at fixed $\{J_1,J_2,J_d\}$ is explained in Appendix \ref{APP-ChiAB}.
Unfortunately, the remaining parameters cannot be obtained from a minimization algorithm because $Q_\X$ is not continuous (the number of PPAs depends on the $J$'s).
On the other hand, as the  number of parameters is reduced, the quality function $Q_\X$ can be evaluated on grids, and after some {trials}, the main minima are eventually found.

The pure kagome model M100 is compatible with the experimental data with ferromagnetic $J_1\sim-12 $\,K, $A=1.037$, $B=-1.2\times10^{-4}\,{\rm K}^{-1}$ but only for $T>70$\,K.

\begin{figure}
\begin{center}
\includegraphics[width=0.49\textwidth]{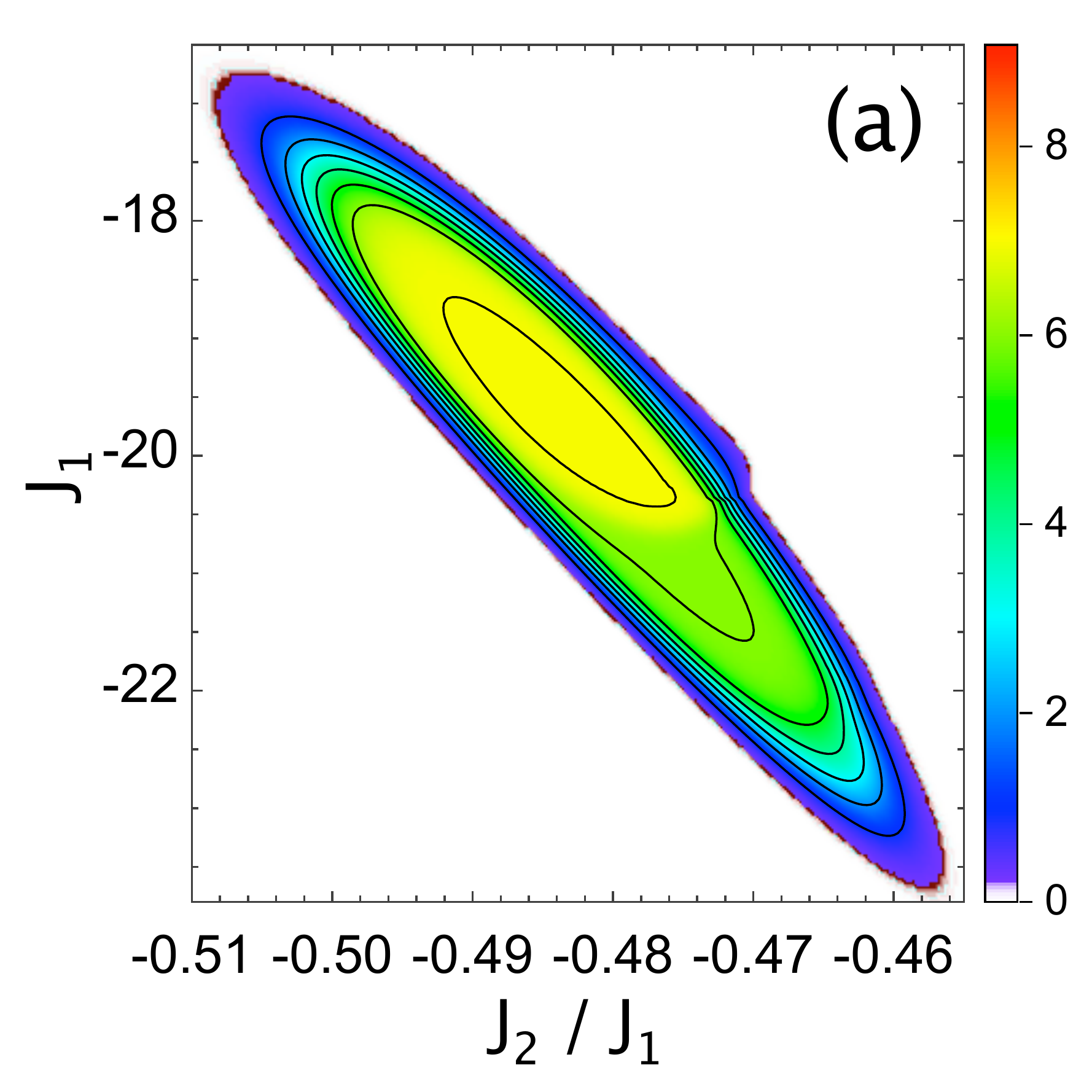}
\includegraphics[width=0.49\textwidth]{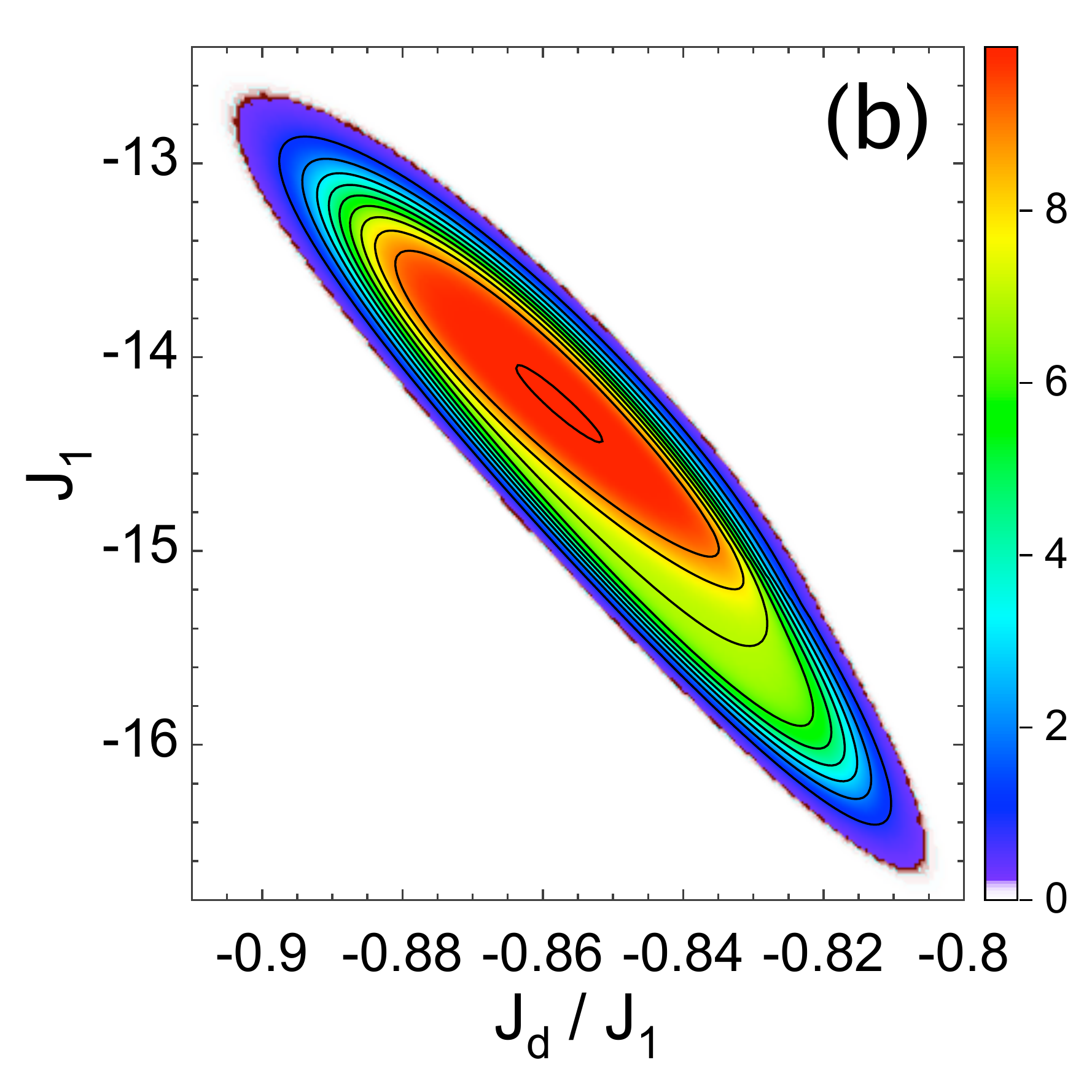}
\includegraphics[width=0.49\textwidth]{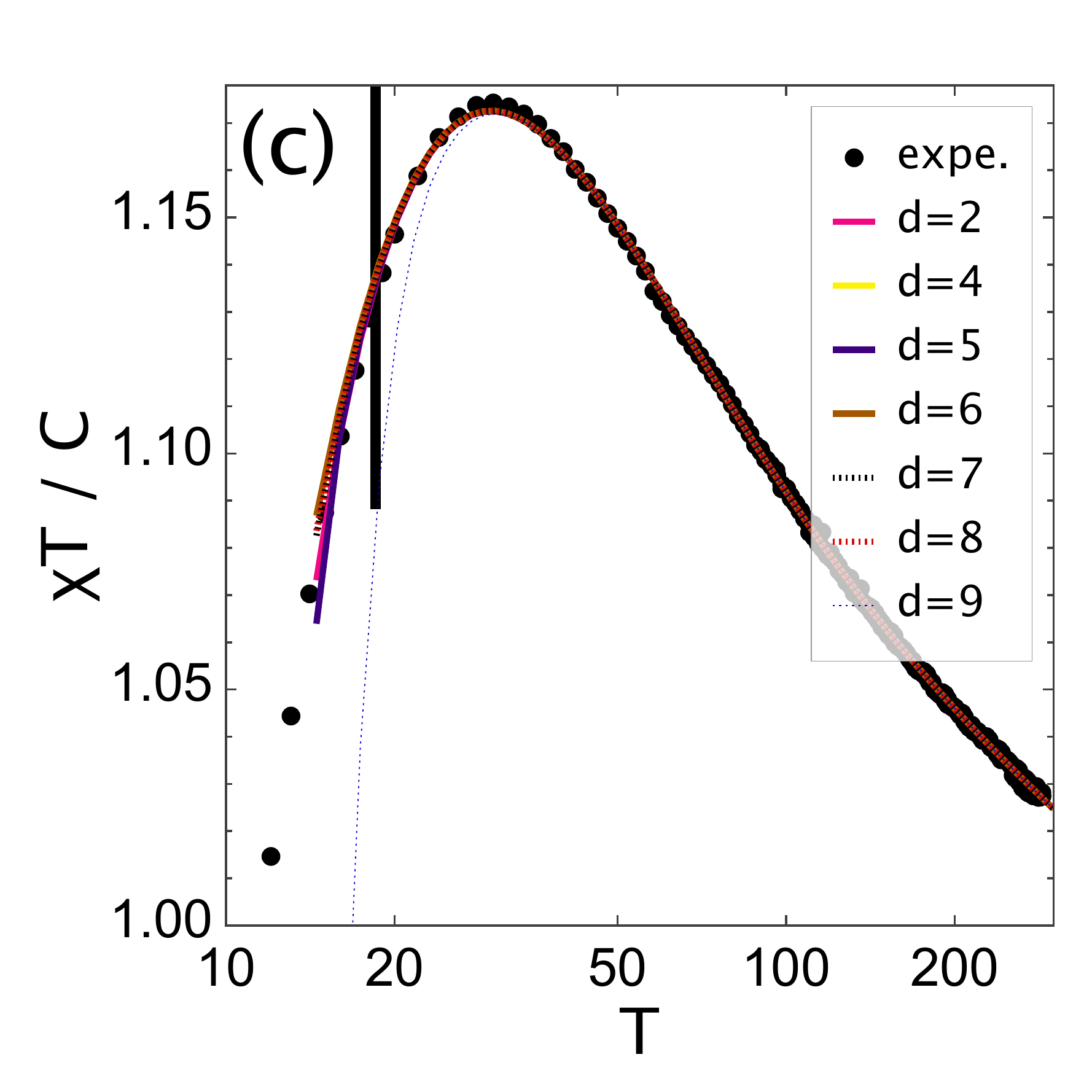}
\includegraphics[width=0.49\textwidth]{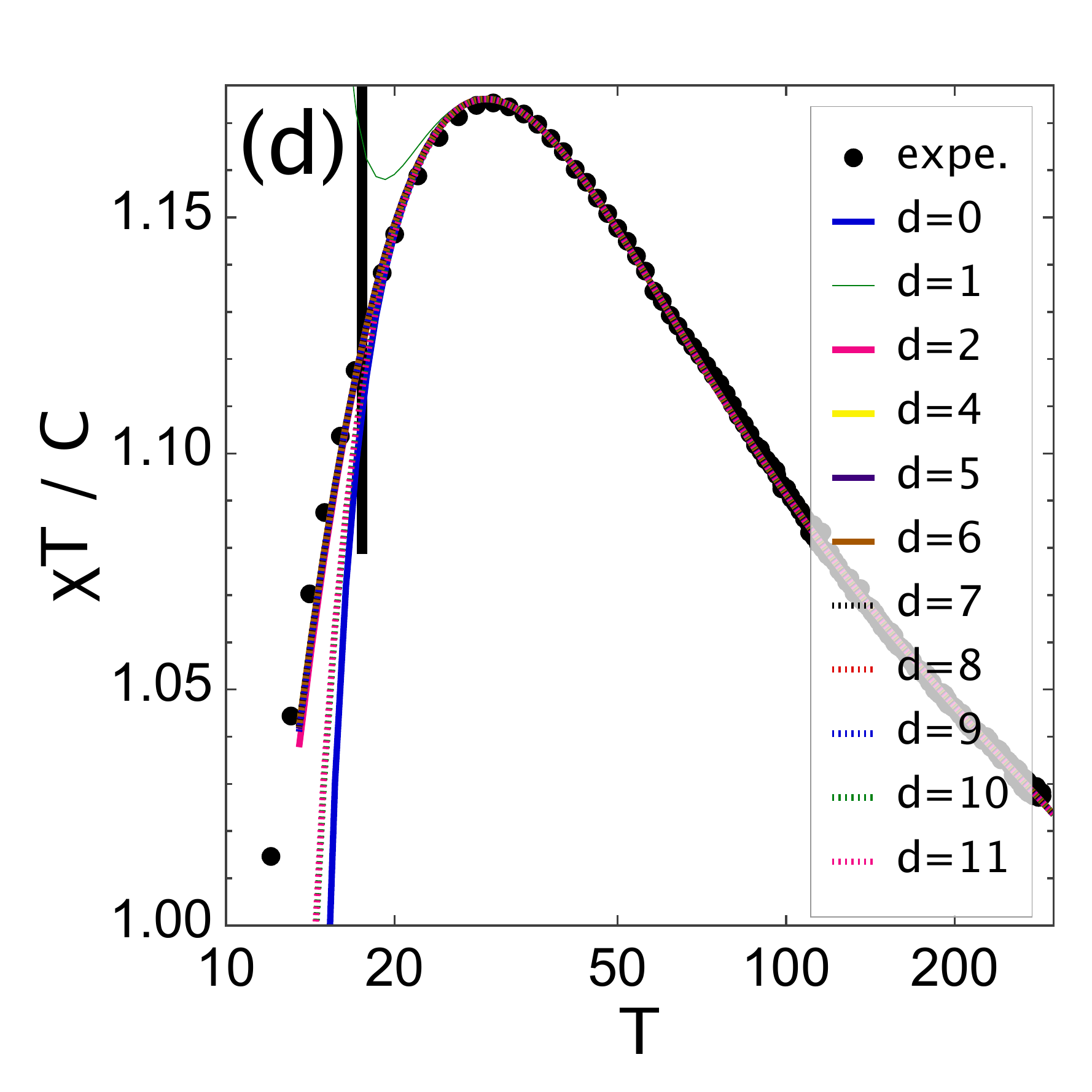}
\caption{(Color online) (a) and (b) Fit quality $Q_\X$ as defined by Eq. (\ref{EQ-Q-def}) with $\epsilon=0.0015$ for the two models (a) $J_1$-$J_2$ and (b) $J_1$-$J_d$ with $T_\min=18.5$ and  $T_\min=17.5$\,K, respectively.
Contour levels are at integer values.
The color scale allows the direct comparison between different models: It is proportional to $Q_\X/n$, where $n$ is the HT-series order.
The best fits are for the highest value of $Q_\X$ (thus, in red).
The parameters at the best points of (a) and (b) are given in the Supplemental Material.\cite{SuppMat}
(c) and (d) Comparison with experiment for these best points of (a) and (b), respectively.
All PPAs at order (c) $n=10$ (d) $n=11$ are shown, and $d$ indicates the degree of the denominator for each PPA (see Appendix \ref{APP-PPA} for the PPA's definition).
Only good PPAs, thick  lines in (c) and (d), are used to compute $Q_\X$ in (a) and (b), respectively.
The thick vertical line indicates $T_\min$.
}
\label{FIG-Chi-Q-COMP12}
\end{center}
\end{figure}

Then, we study models  M120 and  M10d.
Figures.~\ref{FIG-Chi-Q-COMP12}(a) and \ref{FIG-Chi-Q-COMP12}(b) show $Q_\X$  for models M120 and M10d, respectively, whereas Figs.~\ref{FIG-Chi-Q-COMP12}(c) and \ref{FIG-Chi-Q-COMP12}(d)  show all PPAs at the best points of (a) and (b), respectively. 
Note that, in the present method, $Q_\X$ goes rapidly from 0 to some plateau. 
The size of the plateau determines the uncertainties of the fits and depends directly on $\epsilon$.
The fits are of better quality for model M10d with a lower $T_\min$. This is not because the series is known at a higher order but because M10d leads to a better fit of the experimental data around the maximum of $\X T$.
Note that, for these two models, $J_1$ is ferromagnetic whereas $J_2$ and $J_d$ are antiferromagnetic.
In both cases, the precision on $J_2/J_1$ and $J_d/J_1$ is an order of magnitude better than that on $J_1$.

\begin{figure}
\begin{center}
\includegraphics[width=0.75\textwidth]{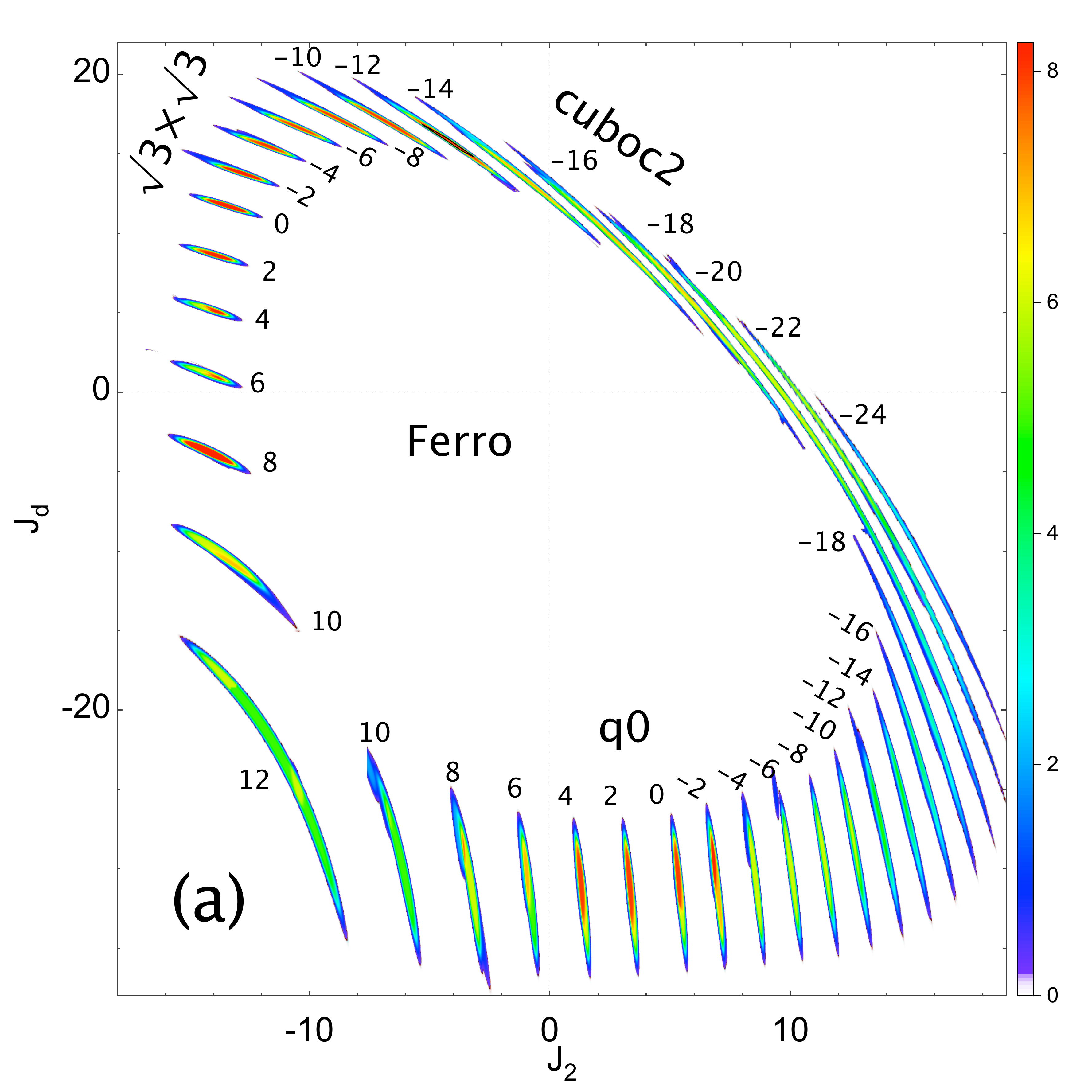}
\includegraphics[width=.49\textwidth]{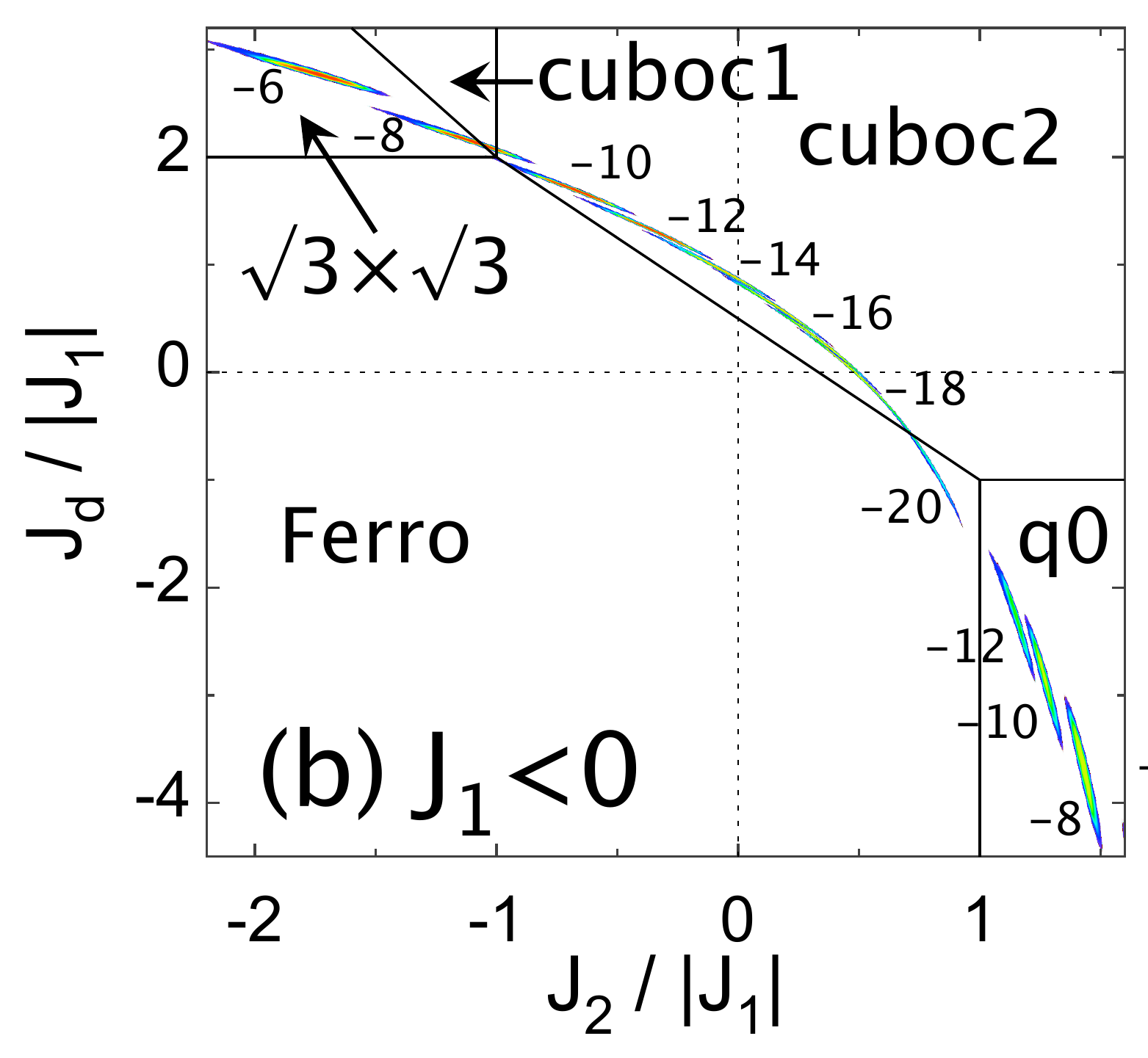}
\includegraphics[width=.49\textwidth]{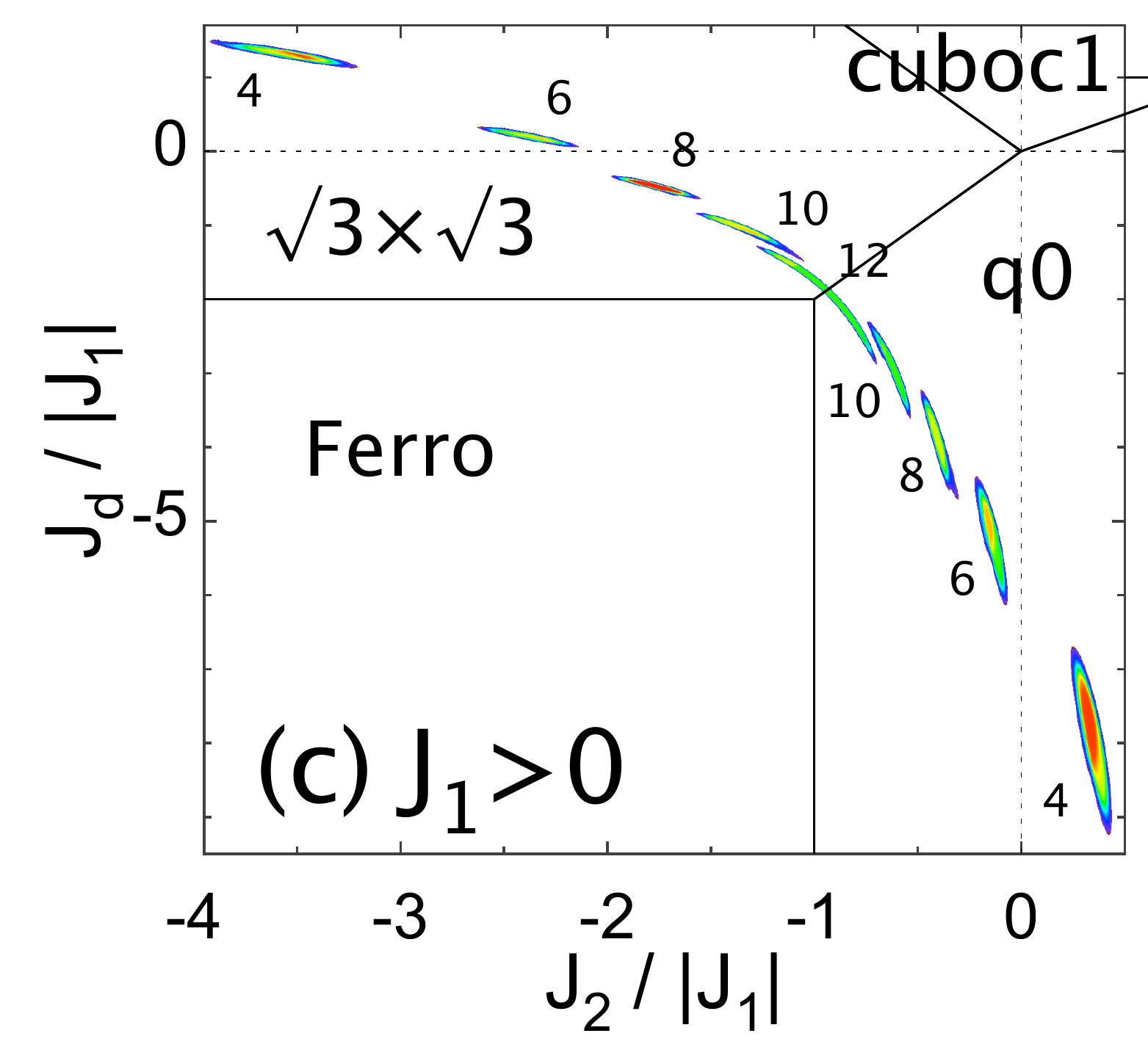}
\caption{(Color online)
Regions of highest-quality fits of the susceptibility (same color code as in Fig. \ref{FIG-Chi-Q-COMP12},  $T_\min=17.5$\,K and $\epsilon=0.002$).
Various cuts of these regions are displayed at fixed $J_1$ (see text). Numbers near each cut indicate the $J_1$ value. 
Symbols [(only indicative in (a)] describe the nature of the order parameter of the classical phase in the corresponding range of parameters.\cite{MESSIO}
(a) gives a global view of the results of $Q_\X$. 
(b) precisely locates these regions of high-quality fits in the classical phase diagrams for ferromagnetic $J_1$
and (c) for antiferromagnetic $J_1$.
The parameters at the best points of each cut of (a) are given in the Supplemental Material\cite{SuppMat}.
In (a), the black parallelogram on the cut-$J_1=-12$ K, visible by zooming it, summarizes the uncertainties on the final best point found at the end of Sec. \ref{SEC-CV}.
}
\label{FIG-ALLQ6}
\end{center}
\end{figure}

With the full model (M12d), we have looked for the solutions at fixed $J_1$ between $-30$ and 30 K. 
We often find two domains of high $Q_\X$.
In a three-dimensional plot of $Q_\X$ versus $J_1$, $J_2$, and $J_d$, the domains of high quality fits (say $Q_\X>6$) fall into a strongly squeezed torus with $J_1$ between $-24$ and 12 K ($Q_\X\sim0$ for $J_1$ outside this interval).
Cuts of these domains at fixed $J_1$  are shown in Fig. \ref{FIG-ALLQ6}(a).
Note that, despite the lower order of the M12d-HT series, these results agree very well with those of models M120 and M10d (Fig.\ref{FIG-Chi-Q-COMP12}).
The sets of optimal parameters are plotted on the classical phase diagram of the \JJJ model\cite{MESSIO} for ferromagnetic and antiferromagnetic $J_1$ in Figs. \ref{FIG-ALLQ6}(b) and \ref{FIG-ALLQ6}(c), respectively.
The best fits appear in various phases of the classical phase diagram nearby the ferromagnetic phase but never in the ferromagnetic phase itself. 
As quantum fluctuations stabilize antiferromagnetic phases and do not change the energy of the ferromagnet, the ferromagnetic phase of the quantum model is expected to have a smaller extent than the classical one, and we are fully confident that all solutions found here fall in an antiferromagnetic quantum phase.
But $\X$ alone is insufficient to determine in which antiferromagnetic phase kapellasite is.

We finish this section with comments on the two parameters $A$ and $B$.
The quantity $A - 1$ in Eq. (\ref{EQ-defZchi}), which measures the uncertainty on $C$ takes values on the order of a few percent in agreement with experimental uncertainties. 
The sum of the Van Vleck and diamagnetic contributions to susceptibility is measured by $B$ and is about $-10^{-4} {\rm K}^{-1}$ which is of the order of expected values.

\section{Describing $C_V(T)$}
\label{SEC-CV}
Throughout this paper, the specific heat stands for the dimensionless specific heat per spin [$C_V\equiv C_V/(Nk_B)$].
It has a spin and a phonon  contribution, $C_V^\spin$ and $C_V^\phonon$, respectively.
At high temperatures, the leading term of $C_V^\spin$ decreases as $J_\CV^2/T^2$, where $J_\CV^2$ is a positive quadratic form of the coupling constants $\{J_a\}$, here $J_\CV=[3/8(J_1^2+J_2^2+J_d^2/2)]^{1/2}$, i.e., at least, $\simeq10$\,K according to the results found in the previous section.
On the other hand, at low temperatures, the $C_V^\phonon$ starts as $(T/T_D)^3$, where $T_D$ is a Debye temperature.
When $J_\CV$ is much smaller than $T_D$, say as for helium-3,\cite{REF-HE3} both terms can be handled independently.
A quick analysis of $C_V^\phonon$ reveals that $T_D\sim170$\,K.
Thus, between 10 and $100$\,K, both contributions are mixed together and we will focus on the fit of $C_V$ on the low-temperature data below $10$\,K.

\begin{figure}
\begin{center}
\includegraphics[width=.49\textwidth]{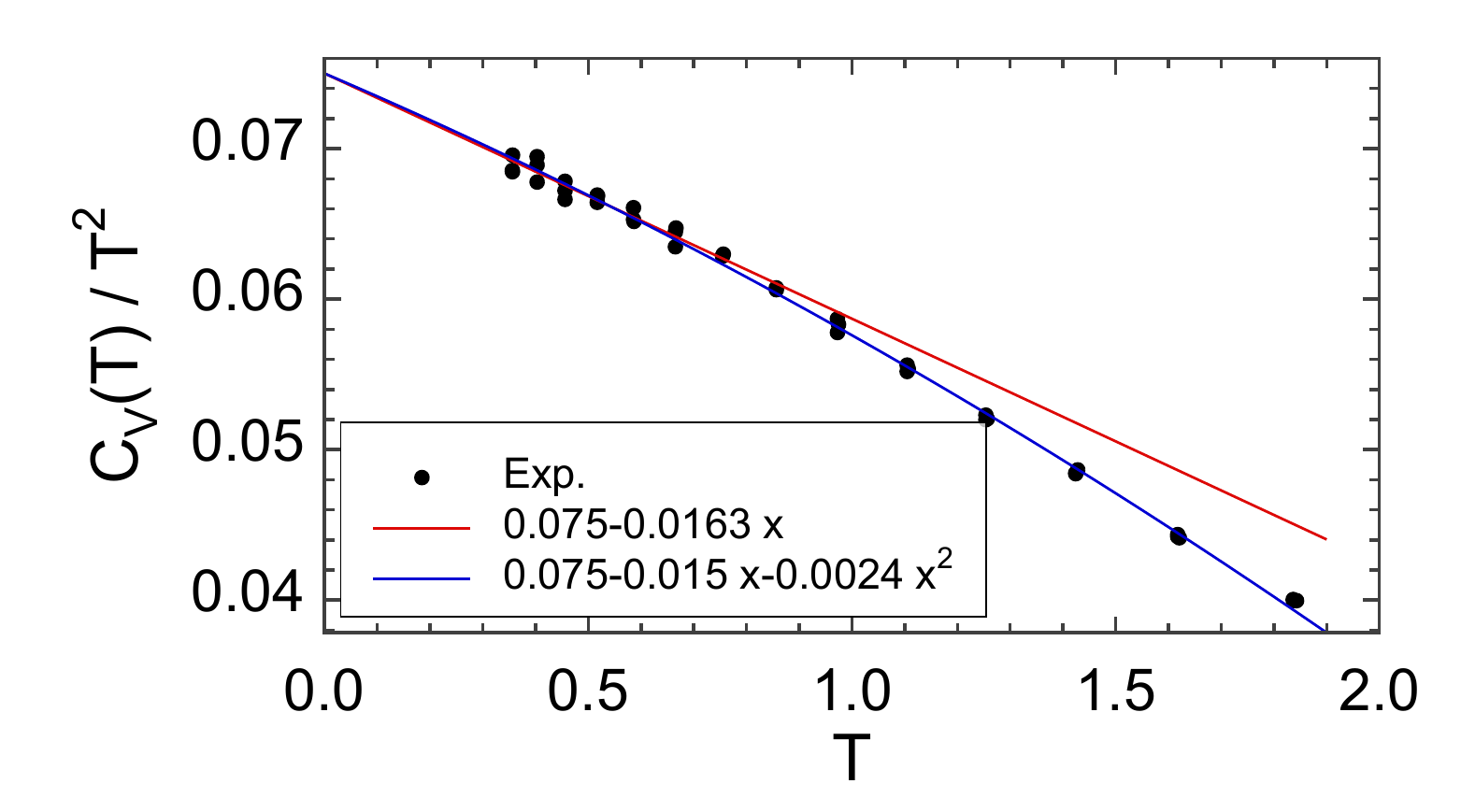}
\caption{(Color online) Low-temperature behavior of the kapellasite specific heat.
Points are experimental data, and the lines are polynomial fits of degree 1 (respectively 2) of the data for $T<0.9$\,K (respectively $T<2$\,K).
}
\label{FIG-CV-LOWT}
\end{center}
\end{figure}

\begin{figure}
\begin{center}
\includegraphics[width=1\textwidth]{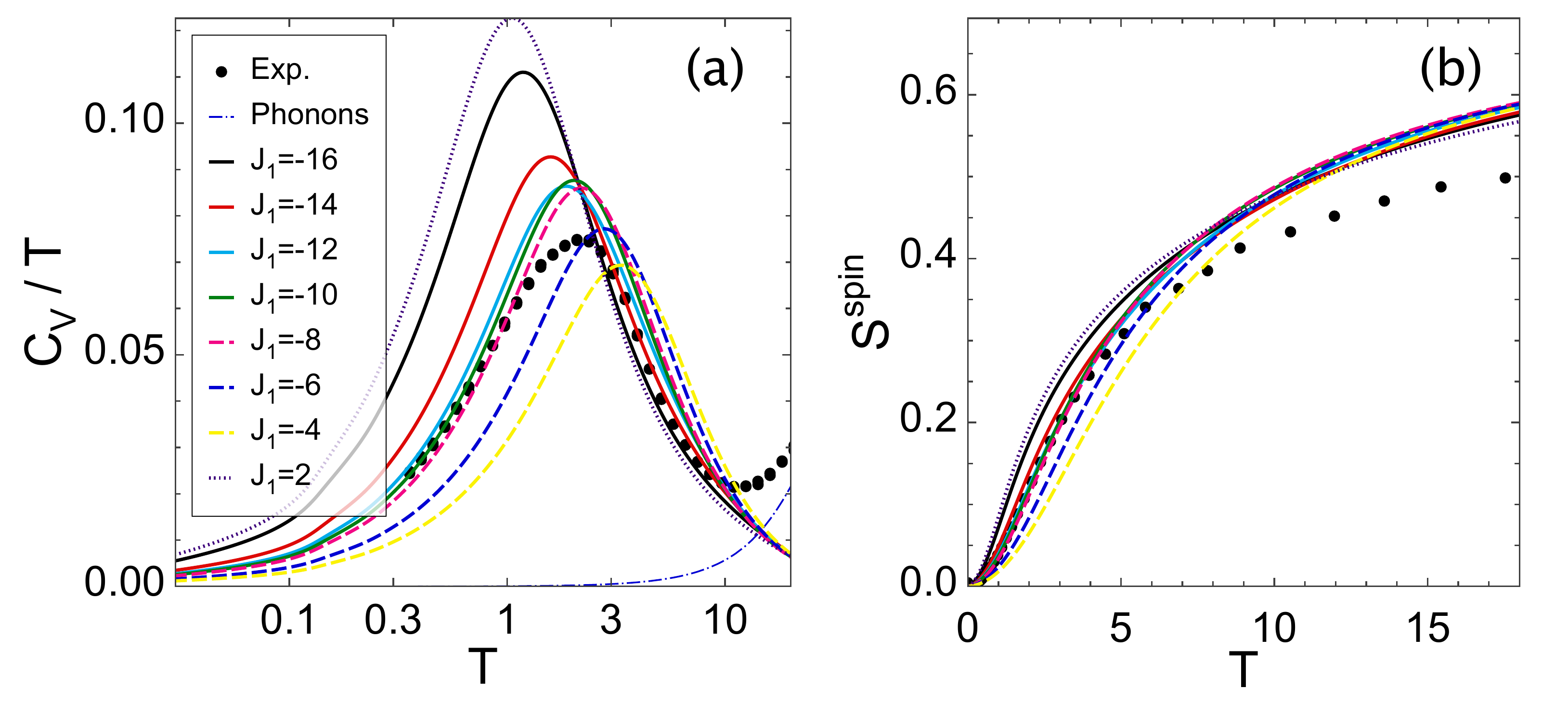}
\caption{(Color online) Comparison with experiments of  (a) $C_V^\spin/T$ and entropy (b) $S^\spin(T)=\int_0^T \!dT\,C_V^\spin/T$ for various parameters of model M12d. 
The dotted line stands for experiments, the dot-dashed line stands for phonons;
full (cuboc2), dotted (\S3), and dashed ($q0$) lines stand for models in the middle of the domains found in the previous sections. The complete sets of parameters are given in the Supplemental Material,\cite{SuppMat} and only the values of $J_1$ are reported in the legend.
In the ``$q0$ domain'' of Fig. \ref{FIG-ALLQ6}, all curves are very similar so only one has been kept in this plot.
In the ``\S3 domain," for $J_1>2$, most of the curves (not shown here) continue to shift to higher temperatures.
}
\label{FIG-CVCOMP}
\end{center}
\end{figure}

In Fig. \ref{FIG-CV-LOWT} we show that fits of $C_V/T^2$, at the lowest available temperature, approach to a constant $0.075(3)\,{\rm K}^{-2}$, compatible with a 2D-antiferromagnetic ground state and the solutions found in the previous section.
Assuming this fit extrapolates to $T=0$ and following a method based on sum rules\cite{KAG-HT,CV-PADE} (see Appendix \ref{APP-CV-PADE}), we calculate $C_V^\spin/T$ per spin for the various models found in the previous section and compare it to experimental data in Fig. \ref{FIG-CVCOMP}.
If none of these models agrees exactly with the experimental data, only those corresponding to $J_1$ between $-14$ and $-6$ K have a maximum at the right position.
So, the position of the $C_V/T$ downturn definitively excludes correlations of the $q0$ type and favors the cuboc2 type described in Messio {\it et al.}\cite{MESSIO}

But, the inset of Fig. \ref{FIG-CVCOMP} also shows that some entropy is clearly missing in all cases above 10 K.
We interpret the data as follows: A large percentage of the spins ($\sim 87\%$, see below) is described by a pure model \JJJ \ below $\sim 3$\,K  whereas  the remaining ones are supposed to be frozen in this low-temperature range  and account for the missing entropy at larger temperatures (5$-$50 K).
Assuming the phonons and this non described part  are negligible at low $T$,
we, thus, set $C_V^\exp=C_V^\spin$ below $T=3$\,K, where
$C_V^\spin$ is represented with PPA, $C_{V,\PPA}^\spin$, as explained in Appendix \ref{APP-CV-PADE}.

Experimental data are given as a list of points $\{T_k,C_{V,k}^\exp\}$.
As in the previous section, we introduce a quality factor $Q_{C_V}$ as:
\begin{eqnarray}
\label{EQ-Q-CV}
	Q_{C_V}&=&\sum_{\{\PPA\}} \M(Z_{C_V,\PPA}^{})\\
\label{EQ-Q-CV-R}
	Z_{C_V,\PPA}^{}&=&\frac1{N_T}\sum_{T_k<3K} \left[ \frac{DC_{V,\PPA}^\spin(T_k)-C_V^\exp(T_k)}{\epsilon T_k}\right]^2\quad
\end{eqnarray}
where $\M$ is a measure function [see Eq. (\ref{EQ-def-M})], $\epsilon=0.0025$ is the uncertainty on $C_V/T$, and $N_T$ is the number of experimental points in the sum.
The parameter $D$ accounts for both mass uncertainty and possible missing entropy and is evaluated as explained in Appendix \ref{APP-ChiAB}.

\begin{figure}
\begin{center}
\includegraphics[width=.235\textwidth]{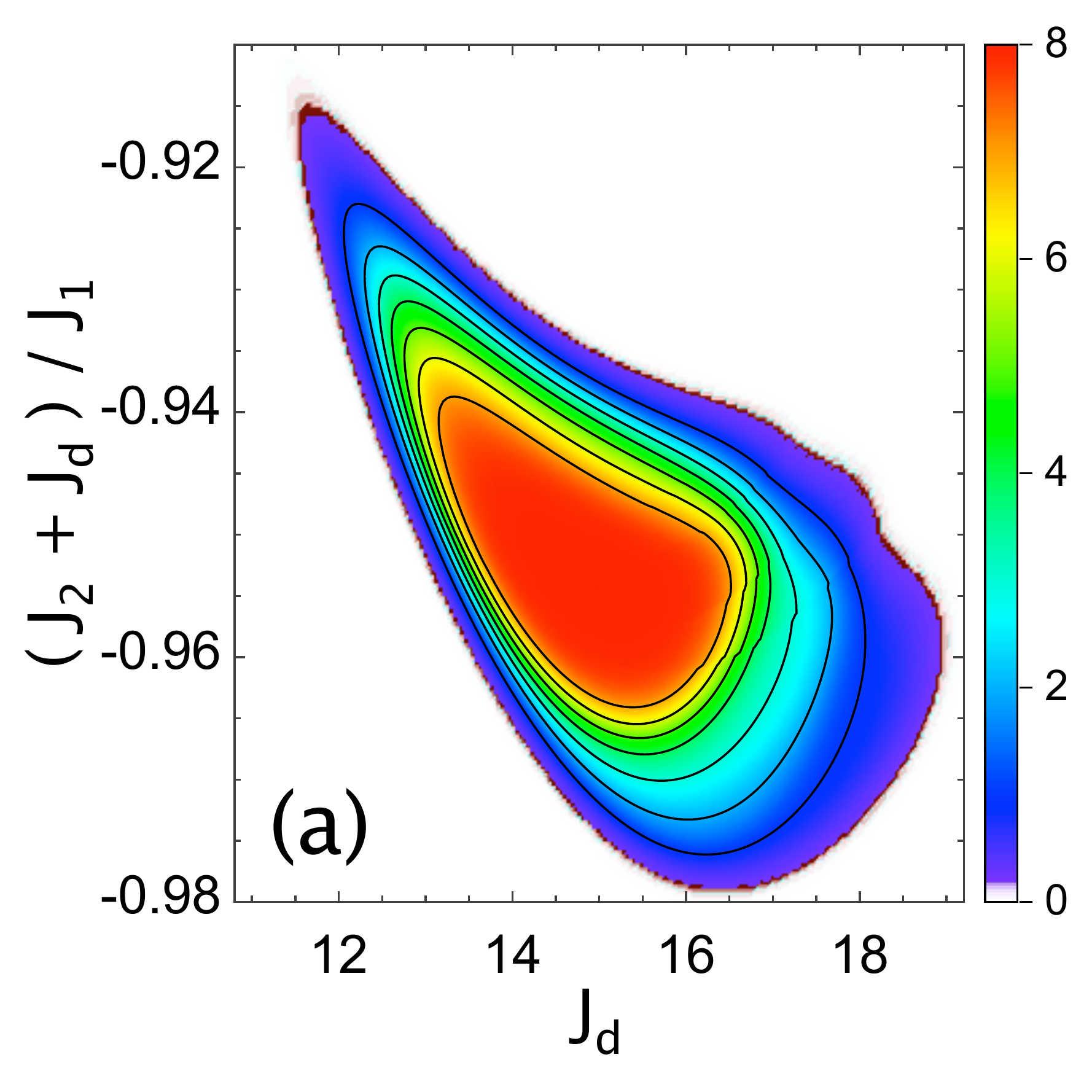}
\includegraphics[width=.235\textwidth]{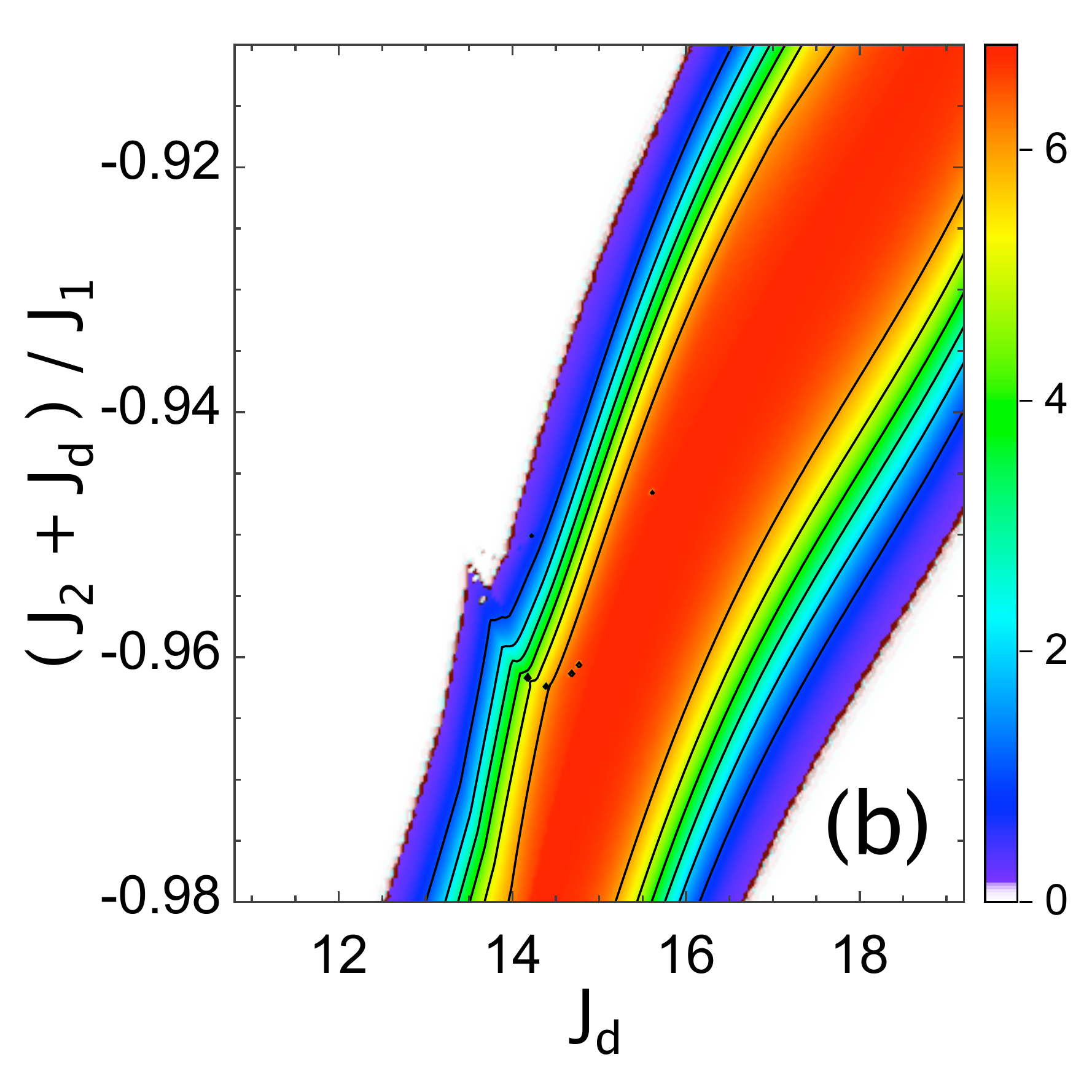}
\includegraphics[width=.235\textwidth]{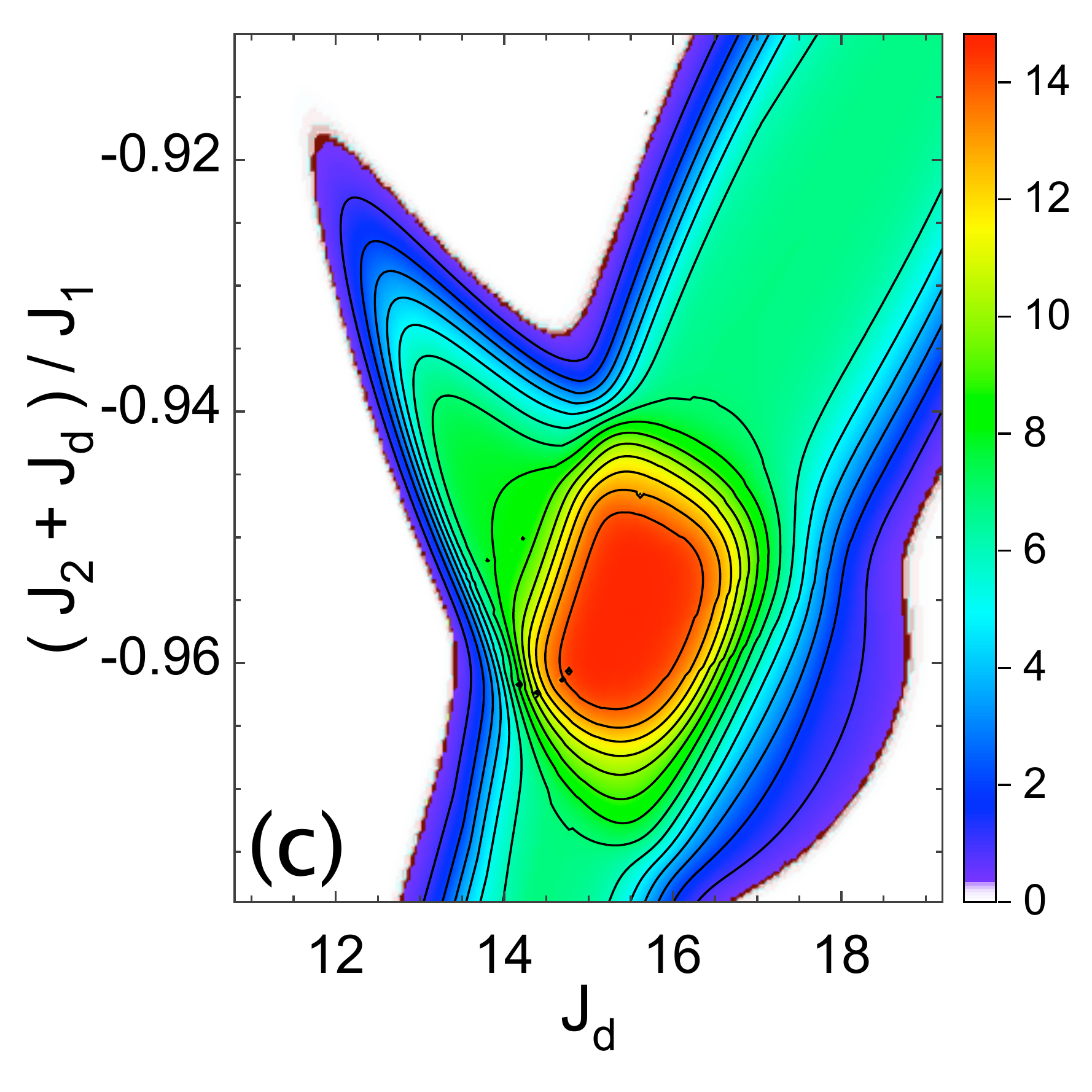}
\includegraphics[width=.235\textwidth]{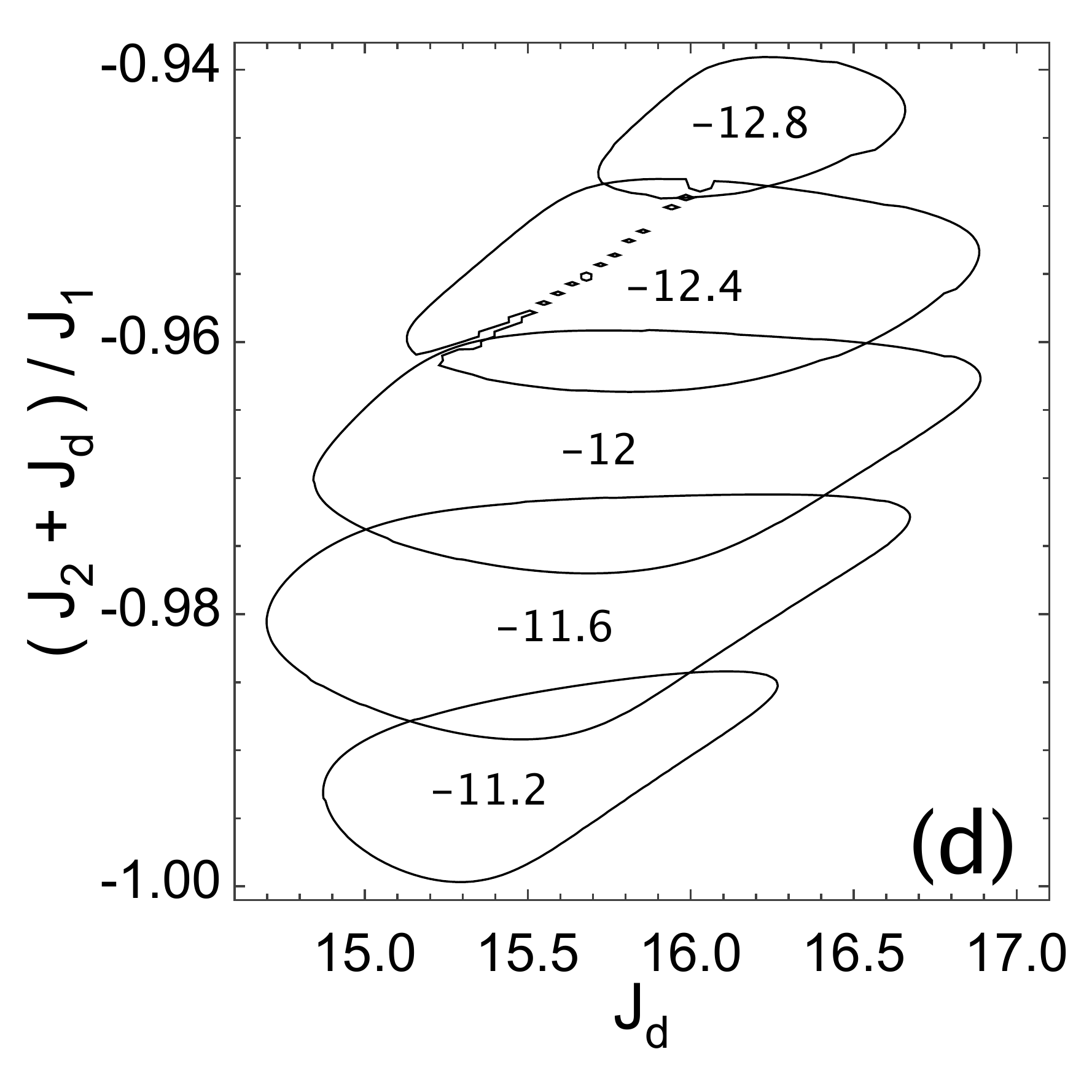}
\caption{(Color online) 
Quality factors (a) $Q_\X$ with $T_\min=17.5$\,K and  $\epsilon=0.0015$, (b) $Q_{C_V}$ and (c) $Q_\X+Q_{C_V}$ at $J_1=-12.4 $\,K versus $J_d$ and $(J_d+J_2)/J_1$.
Contours are at integer values.
Note that the range of $(J_2+J_d)/J_1$ is 2 orders of magnitude less than that of $J_1$ and $J_d$.
The quality factors increase rapidly from almost 0 to a maximum value represented by plateaus.
(d) The lines are the contours $Q_\X+Q_{C_V}=14$ for various values of $J_1$ indicated by the numbers near each line.
In a three-dimensional plot, $Q_\X+Q_{C_V}>14$ is contained roughly in a tube that ends abruptly at $J_1>-11$ and in a small cone at $J_1\sim-13$\,K. This can be understood by viewing the band of high $Q_{C_V}$ (b) entering the domain of $Q_\X>0$  (a) from the right at $J_1\sim-10 $\,K and gliding to the left as $J_1$ decreases.
}
\label{FIG-QChi-QCV}
\end{center}
\end{figure}

$C_{V,\PPA}^\spin$ depends on the unknown ground state energy per spin $e_0$.\cite{Note-Ferro}
Appendix \ref{APP-e0} describes how $e_0$ is evaluated using an another quality factor.
As a consequence, computing $Q_{C_V}$ is much more demanding and less stable and the figures $Q_{C_V}(J_1,J_2,J_d)$ present several spurious discontinuities. 
In the domain of interest, keeping the good PPAs to compute $e_0$ and $Q_{C_V}$ removes most of these discontinuities. 
Figure \ref{FIG-QChi-QCV} shows, at $J_1=-12.4 $\,K, the results for $Q_\X$, $Q_{C_V}$ and $Q_\X+Q_{C_V}$.
The choice of axis, $J_d/J_1$ and $(J_2+J_d)/J_1$, replaces the strongly squeezed domain of high $Q_\X$ (see Fig.\ref{FIG-ALLQ6}) into a more compact one.
The high-$Q$ domains are different for $\X$ and $C_V$ and may eventually overlap as shown in Fig. \ref{FIG-QChi-QCV}(c).
Choosing a threshold for $Q_\X+Q_{C_V}$ determines the domain of validity of the overall fit.
The plateaus around the maxima being surrounded by sharp walls, the determination of the best-parameter range is rather independent of the threshold.
Repeating the process for various $J_1$'s, in Figure \ref{FIG-QChi-QCV}(d), we show the overall constraints on the parameters where the best fits of $\X$ and $C_V$ are found.
This figure shows that two parameters, say $J_1$ and $J_d$, are defined with a larger uncertainty than the rather well defined ratio $(J_2+J_d)/J_1$.
The results are summarized as:
\begin{eqnarray}
\label{EQ-JJJFINAL}
\nonumber
J_1&=&-12.0(8)\quad\INK\\
J_d&=&15.6(9)+0.5\,(J_1+12)\quad\INK\\
\nonumber
\frac{J_2+J_d}{J_1}&=&-0.97(1)-0.03\, (J_1+12)
\end{eqnarray}

\begin{figure}[t]
\begin{center}
\includegraphics[width=0.49\textwidth]{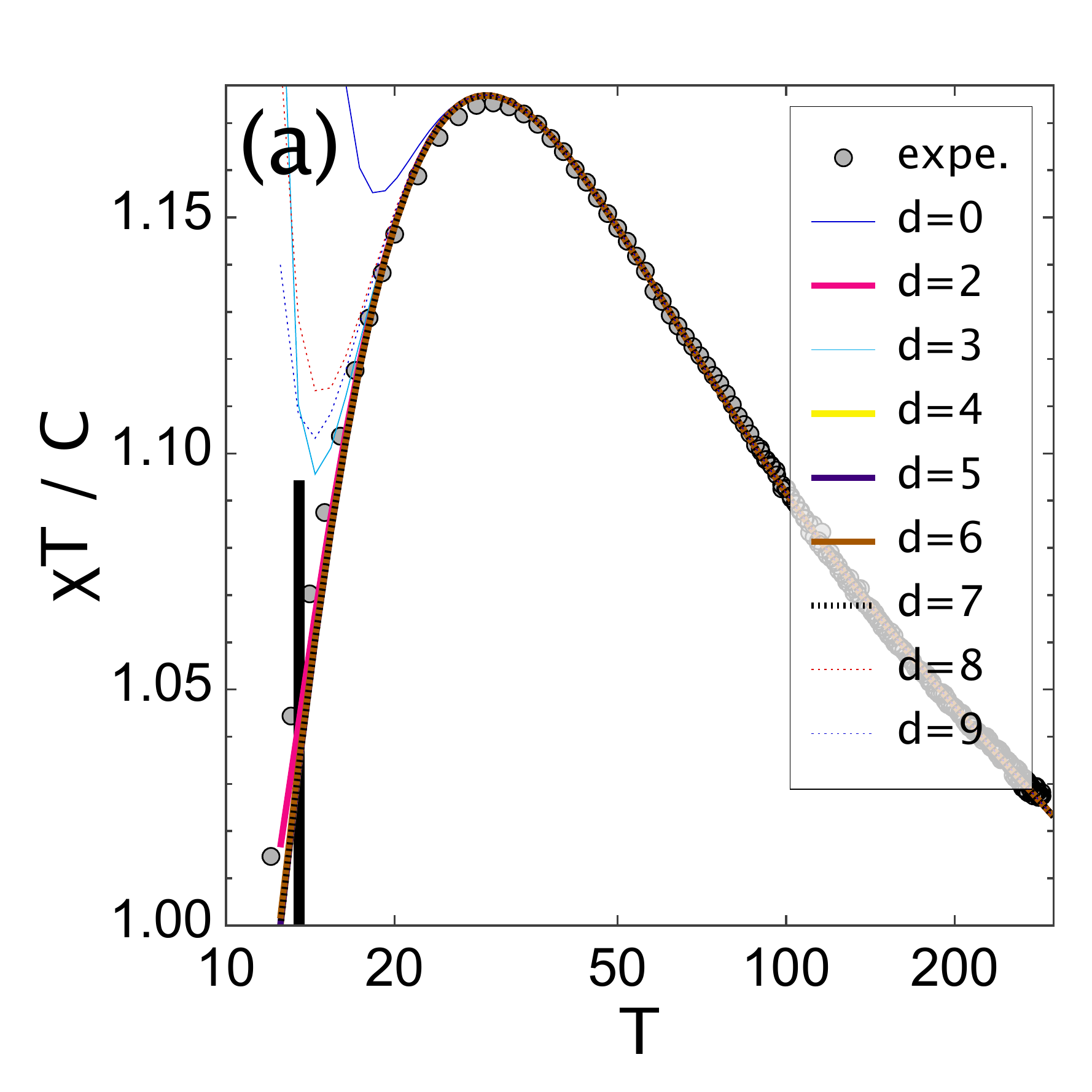}
\includegraphics[width=0.49\textwidth]{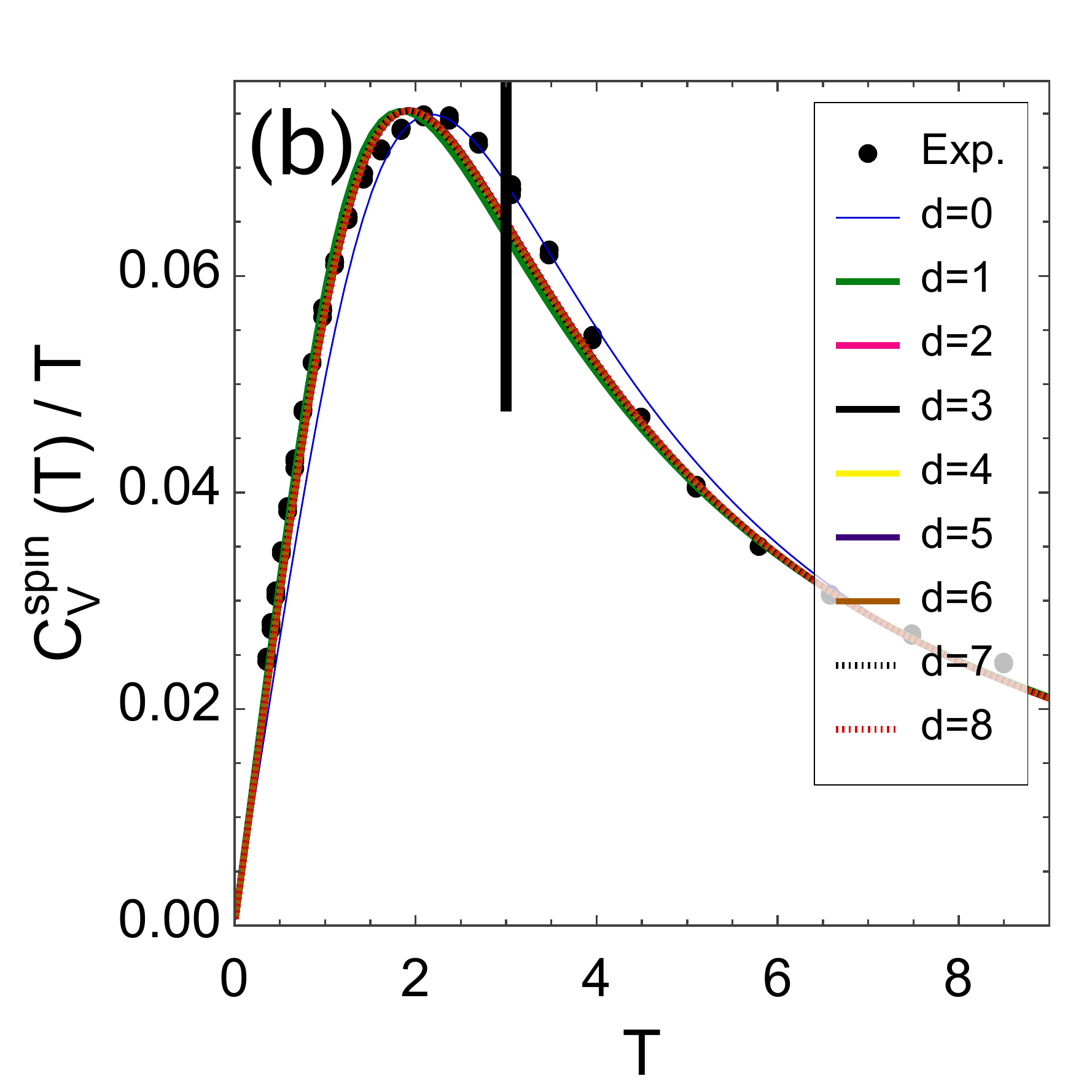}
\caption{(Color online)
Comparison with experiments for $J_1=-12$, $J_2=-4$ and $J_d=15.6$\,K.
(a) Magnetic susceptibility with $A=1.027$, $B=-10^{-4}$\,K$^{-1}$, and $T_\min=16.5$\,K [thick vertical line, see Eq. (\ref{EQ-defZchi})].
(b) Specific heat with $D=0.863$ and $e_0=-15.674$\,K.
The vertical line stands for $T_\max$ [see Eq. (\ref{EQ-Q-CV-R})].
}
\label{FIG-FINAL}
\end{center}
\end{figure}

Figure \ref{FIG-FINAL} shows the comparison of both $\X T$ and $C_V/T$ at the center of the best domain, i.e., $J_1=-12$, $J_d=15.6$, and $J_2=-4$\,K. The uncertainties on $\{J_2,J_d\}$ are well represented by the red part of the cut at $J_1=-12$\,K of Fig. \ref{FIG-ALLQ6}(a) or \ref{FIG-ALLQ6}(b), visible by zooming it.

\section{Conclusion}
\label{SEC-Conclusion}

We have fitted the spin contribution of the magnetic susceptibility and specific heat  experimental data with a spin-1/2 \JJJ Heisenberg model on the kagome lattice (see Fig. \ref{FigLattice}).
In contradiction to the \textit{ab initio} calculations of Janson {\it et al.},\cite{Janson} the nearest-neighbor coupling is ferromagnetic. This is at variance with herbertsmithite where the nearest neighbor interaction is strong and antiferromagnetic.
This can be traced back to the Cu-$\mu_3$OH-Cu-bonding angle being $\sim13^\circ$ smaller in kapellasite.\cite{kapella2,wills2008}
The isostructural compound Haydeeite Cu$_3$Mg(OH)$_6$Cl$_2$, also has a ferromagnetic first neighbor interaction but is in the  ferromagnetic domain. \cite{Haydeeite,kapella2}
This is not the case for kapellasite where the $J_2$ and $J_d$ exchange couplings compete to form a non magnetic compound.

The spin susceptibility is relatively easy to reproduce and imposes strong correlations of $J_1$, $J_2$ and $J_d$.
All solutions stay in  anti-ferromagnetic  domains of the classical phase diagram, but different phases remain potential candidates.\cite{MESSIO}
The main distinctive features of the specific heat data are the low-$T$ downturn in $C_V/T$ at about 2\ K, characteristic of the competitive exchange couplings and a clear $T^2$ dependency excluding a ferromagnetic ground state. The peak strongly constrains the parameters.
The best domain for \textit{both the magnetic susceptibility and the specific heat } is obtained for
 $J_1\sim-12$\,K,  a small ferromagnetic $J_2\sim-4$\,K  and a large antiferromagnetic $J_d \sim 15.6$\,K [Eq. (\ref{EQ-JJJFINAL})].
These  parameters predict the system to have cuboc2 correlations as found independently by neutron-scattering experiments.\cite{CUBOC2PRL}
These competitive exchange energies give a ferromagnetic behavior of the magnetic susceptibility at high temperatures and an antiferromagnetic one at low temperatures.

However, the agreement between experiment and theory is not yet as good as a quick glance at Fig. \ref{FIG-FINAL} would suggest.  There is about 14\% of missing entropy in our description [$D=0.863$ in Eq. (\ref{EQ-Q-CV-R}), whereas, the mass uncertainty is of only about a few percent].
As it is improbable that this missing entropy will be found at ultralow temperatures below our present measurements, it must be released at intermediate temperatures between 5 and 20 K, where we have not succeeded to fit the full specific heat variations with this spin model and phonon contributions.

Disorder might be invoked to explain this difficulty.  
In fact, the actual chemical formula of the synthesized compound, determined with neutron powder diffraction,\cite{kapella2} is (Cu$_{0.73}$Zn$_{0.27}$)$_3$(Zn$_{0.88}$Cu$_{0.12}$)(OH)$_6$Cl$_2$ with 27\% Zn on the Cu sites of the kagome lattice and 12\% Cu  on the ‘‘hexagonal’’ Zn site.
In the classical model, this concentration of Zn on the kagome sites is not enough to kill the long-range  cuboc2 correlations,  the threshold being at about 40\%.\cite{Brieuc} The pure quantum model is certainly softer, and the nature of its ground state is still an open question. 
Heuristically, the presence of vacancies or extra spins can induce \textit{a priori} two phenomena: either the manifestation of weakly coupled local spin oscillations (the so-called ``free spins") or the freezing of singlets. The``free impurity spins" would show up in spin susceptibility  in  differences between the bulk SQUID measurements of  the magnetization and the local NMR data. As no such phenomenon has been observed in this compound in the range of the present experiments, we do not believe that it would help in understanding the results of the fits. The second possible (quantum) phenomena is a partial freezing of  isolated singlets   along  diagonal $J_d$ bonds (recall that $J_d$ is the largest antiferromagnetic energy in this compound $\sim 15.6$\,K). These singlets would not show up  in the specific heat at temperatures lower than a fraction of $J_d$,  explaining the $D$ constant $\sim 0.87$  needed to fit the specific heat data at temperatures lower than 3\,K.  At higher temperatures, the liberation of these spins, through thermal excitations of the local singlets, would explain that the spin susceptibility measurements and fits above 17\,K  give the correct amount of spins in the sample.
A better description of this phenomenon is out of the scope of the present approach, but could perhaps be explored with exact diagonalizations.

In the present paper, DM interactions have been neglected.
In fact, the lack of an inversion center on the magnetic bounds allows for DM interactions of spin-orbit origin. In the cuprates, these couplings are usually estimated on the order of $1/10$ of the super exchange couplings, and in herbertsmithite, they were measured on the order of a few percent.\cite{Zorko2008}
In herbertsmithite, the influence of this small coupling is emphasized by the presence of a nearby quantum critical point.\cite{helton07,cepas2008,messioDM2010} 
The situation in kapellasite is quite different:  Whereas neutron scattering in herbertsmithite is  essentially featureless\cite{Helton2010,Han12nature} the experimental evidence of short-range cuboc2 correlations in kapellasite is strong,\cite{CUBOC2PRL} and the results of the present analysis independently point to the same conclusion: The $J_2$ and $J_d$ parameters locate the system in the ``cuboc2 domain," far away from any critical point (the cuboc2 ferro transition is a strong first order transition\cite{Domenge2005}). Extending the present fit to take Dzyaloshinskii-Moriya interactions into account would slightly change the exchange parameters but would not move the system away from the present phase. With these caveats in mind, the present model is the best-effective model that we are able to build.

Acknowledments: We are greatly indebted to our collaborators, first of all, R. Colman and A. Wills who synthesized this compound and brought it to our attention.\cite{kapella1,kapella2} The inelastic neutron-scattering data of B. F{\aa}k {\it et al.} was the second considerable step in catching our attention for kapellasite. Extensive discussions with them during the two years of this collaboration have been invaluable. C.L. acknowledges very interesting discussions with O. Janson.

\appendix

\section{PPA}
\label{APP-PPA}
For fixed values of the coupling constants $J_1$, $J_2$ and $J_d$, we evaluate the order-$n$ HT polynomial $P_n(x)$, around $x=0$, of the magnetic susceptibility $\X^\HT_n(\beta)$ or of the specific heat $C_{V,n}^\spinHT(e)$ where $e$ is the energy per spin.
From a polynomial $P_n(x)$ of degree $n$, we calculate the $(n+1)$-rational fractions $N_{n-d}(x)/D_d(x)$, having the same series as $P_n(x)$ around $x=0$, with the degree $d$ of $D$ running from 0 to $n$: They are the so-called Pad\'e approximants of $P_n(x)$. From this list, we discard all the Pad\'e approximants which have zeros
 either in $N$ or $D$ in the whole interval of variation of $x$, i.e., in $[0,\infty]$ for $\X^\HT_n(\beta)$ and $[e_0,0]$ for $C_{V,n}^\spinHT(e)$, where $e_0$ is the ground state energy.
The remaining ones are called the PPA.  By varying the coupling parameters, the number of PPAs may eventually change. Thus, all functions built on the sum over the PPAs may be discontinuous. Unfortunately, this prevents using minimization powerful methods. This is the price to pay when using PPAs.

\section{DETERMINATION OF PARAMETERS $A$ AND $B$ OF $Q_X$}
\label{APP-ChiAB}
From Eq. (\ref{EQ-Q-def}) or Eq. (\ref{EQ-Q-CV}), we have
\begin{eqnarray}
	Q_X&=&\sum_{\PPA} \M(Z_{X,\PPA})\\
	Z_{X,\PPA}&=&\frac1{\epsilon^2 N_T}\sum_{k} \left[ A F_{\PPA}(T_k) +BT_k -F_k^\exp \right]^2,\quad
\end{eqnarray}
where $X=\X$ or $C_V$ and $B$ is 0 for $C_V$ and the measure function is defined in Eq. (\ref{EQ-def-M}). 
$F$ stands for $\X T/C$ if $X=\X$ and $C_V/T$ if $X=C_V$. The derivatives of $Q_X$ with respect to $A$ and $B$ are as follows:
\begin{align}
\nonumber
	\frac{\partial Q_X}{\partial A}
\nonumber
		&=\sum_{\{\PPA\}} \M'(Z_{X,\PPA})\frac{\partial Z_{X,\PPA}}{\partial A}\\
\nonumber
		&=\frac{2}{\epsilon^2N_T}\sum_{\{\PPA\}} \M'(Z_{X,\PPA})	\\&\times
\label{EQ-DIFFA}
		\sum_k\left[A F_\PPA(T_k)\!-\!BT_k\!-\!F_k^\exp\right] F_\PPA(T_k)\\
\nonumber
	\frac{\partial Q_X}{\partial B}
	&=\frac{2}{\epsilon^2N_T}\sum_{\{\PPA\}} \M'(Z_{X,\PPA}) \\&\times
\label{EQ-DIFFB}
		\sum_{k=1}^{N_T}\left[A F_\PPA(T_k)-BT_k-F_k^\exp\right] T_k,
\end{align}
where $\M'(x)$ is the derivative of $\M(x)$.
We look for $A$ and $B$ that cancel out these derivatives.
If the weights $\M'(Z_{X,\PPA})$ are independent of $A$ and $B$, these equations are linear and are easily solved.
Assuming the weights are smooth functions of $A$ and $B$, we solve this problem iteratively.
We choose, as initial point, the $A$ and $B$ solutions of the best PPA [highest $\M(Z_{X,\PPA})$],
\begin{align}
	A_\PPA=&\frac1\Delta \left(\overline{T\, F^\exp} \,\overline{T\, F_\PPA}  -\overline{T^2}\,\overline{F_\PPA\, F^\exp}  \right)&\\
	B_\PPA=&\frac1\Delta \left(\overline{F_\PPA\,T}\,\overline{F_\PPA\, F^\exp}  -\overline{T\, F^\exp} \,\overline{ F_\PPA^2} \right)&\\
	\Delta=&\overline{T\, F_\PPA}^2 -\overline{T^2}\,\overline{F_\PPA^2} &
\end{align}
where $\overline X$ means the mean value over the set of temperatures.
This first estimation of $A$ and $B$ is then used to compute the weights $\M'(Z_{X,\PPA})$ in Eqs. (\ref{EQ-DIFFA}) and (\ref{EQ-DIFFB}), and new $A$ and $B$ are given by
\begin{align}
	A=&\frac1\Delta \left(\left<\overline{T\, F^\exp}\right> \left<\overline{T\, F_\PPA}\right>  -\left<\overline{T^2}\right>\left<\overline{F_\PPA\, F^\exp}\right>  \right)&\\
	B=&\frac1\Delta \left(\left<\overline{F_\PPA\,T}\right>\left<\overline{F_\PPA\, F^\exp}\right>  -\left<\overline{T\, F^\exp}\right> \left<\overline{ F_\PPA^2}\right> \right)&\\
	\Delta=&\left<\overline{T\, F_\PPA}^2\right> -\left<\overline{T^2}\right>\left<\overline{F_\PPA^2}\right> &
\end{align}
where  $\left<\overline X\right>$ means the average value over the set of temperatures and over the PPA with the weights $\M'(Z_{X,\PPA})$.
This procedure is iterated until convergence by calculating the new weights at the new $A$ and $B$.
The convergence is quick and a couple of iterations are sufficient for a relative precision of $10^{-5}$ on $A$ and $B$.

\section{PAD\'E APPROXIMANT FOR $\CV$}
\label{APP-CV-PADE}
Here, we recall how to evaluate the specific heat at all temperatures using sum rules.\cite{CV-PADE,KAG-HT}
For Heisenberg models on two-dimensional lattices, as no phase transitions are expected at finite temperatures, the thermodynamic functions are continuous. 
The entropy per spin versus the energy per spin $s(e)$ is more suitable than $C_V(T)$ as it is constrained to start at the ground-state energy $e_0$ with an entropy $s=0$ and end at $e=0$ and $s=\ln2$ at infinite temperatures.
Moreover, this is a monotonic increasing function $\beta=1/T=s'(e)$ with negative curvature $C_V=-s'(e)^2/s''(e)$.

From the HT-series expansion of $\CV(T)=\sum_{i=2}^n a_i \beta^i$ (see Supplemental Material\cite{SuppMat} for the expression of $a_i$ versus $J_1$, $J_2$, $J_d$) with $\beta=1/T$,  we obtain the HT series of $s(T)$ and $e(T)$ as
\begin{align}
\nonumber
	s(T)=&\ln2-\int_T^\infty \!dT'\,\frac{\CVspin(T')}{T'}\\
\label{EQ-defSTHT}
	=&\ln 2 -\sum_{i=2}^n \frac{a_i}i \beta^i+O(\beta^{n+1})\\
\nonumber
	e(T)=&-\int_T^\infty \!dT'\,\CVspin(T')\\
	=& -\sum_{i=2}^{n-1} \frac{a_{i+1}}i \beta^i+O(\beta^{n})
\end{align}
where we use $s(T=\infty)=\ln2$ and $e(T=\infty)=0$.
The HT-series expansion of $s(e)$,
\begin{eqnarray}
\label{EQ-defse}
	s(e)&=&\sum_{i=0}^n b_i e^i
\end{eqnarray}
is obtained order by order.

We assume a low-temperature power law for $\CV(T)$,
\begin{align}
\label{EQ-CVlowT}
 C_V(T)_{T\to0}\simeq (C_0T)^\alpha.
\end{align}
Then, $s(e)\propto (e-e_0)^{1/\mu}$ for $e$ around $e_0$ where $e_0$ is the ground-state energy and $\mu=1+1/\alpha$.
We define an analytic function in the interval $[e_0,0]$,
\begin{eqnarray}
\label{EQ-defGe}
	G(e)=\frac{s(e)^\mu}{e-e_0}.
\end{eqnarray}
The HT-series expansion for $G(e)$ is obtained from
\begin{eqnarray}
\label{EQ-defGeHT}
	G(e)&=&-\frac{(\ln 2)^\mu}{e_0}
	\left[\sum_{i=0}^{n}F_i(\mu)\frac{P(e)^i}{i!}\right]
	\left[\sum_{i=0}^n (e/e_0)^i\right],\\
	P(e)&=&\frac{s(e)}{\ln2}-1=\sum_{i=2}^n\tilde b_i e^i,
\end{eqnarray}
where $\tilde b_i=b_i/\ln2$ ($b_1=0$)  and $F_i(\mu)=\Gamma(\mu+1)/\Gamma(\mu+1-i)=\mu(\mu-1)\cdots(\mu+1-i)$. Keeping only terms up to order $n$ defines $G^\HT(e)$.
Note that $P(e)^i$ starts at order $2i$.

Then, $G^\HT(e)$ is transformed in all possible Pad\'e approximants noted $G^\HT_{d}(e)$ where $n-d$ and $d$ are the numerator and denominator degrees.
We keep only the PPA denoted $G^\HT_{d^*}(e)=N_{n-d^*}(e)/D_{d^*}(e)$ whose numerator and denominator have no zero inside $[e_0,0]$.
The value $G_{d^*}^\HT(e_0)$ is related to $C_0$ [see Eq. (\ref{EQ-CVlowT})] by $G_{d^*}^\HT(e_0)=C_{0,d^*}^{}(\alpha+1)/\alpha_{}^\mu$.

From $G_{d^*}^\HT$, we obtain $s(e)$, its first derivatives,
\begin{align}
\label{EQ-defdse}
	s_{d^*}(e)=&\left[(e-e_0)\frac{N(e)}{D(e)}\right]^{1/\mu}\\
\label{EQ-defdspe}
	\mu\frac{s_{d^*}'(e)}{s_{d^*}(e)}=&\frac1{e-e_0}+\frac{N'(e)}{N(e)}-\frac{D'(e)}{D(e)}\\
	\nonumber
	 \mu\frac{s_{d^*}''(e)}{s_{d^*}(e)}=&\mu\left[\frac{s_{d^*}'(e)}{s_{d^*}(e)}\right]^2+\frac{N''(e)}{N(e)}-\frac{D''(e)}{D(e)}\\
\label{EQ-defdsse}
	&-\left[\frac{N'(e)}{N(e)}\right]^2+\left[\frac{D'(e)}{D(e)}\right]^2-\frac1{(e-e_0)^2}
\end{align}
Then we deduce $\beta(e)=1/T(e)=s_{d^*}'(e)$ and $\CVspin(e)=-[s_{d^*}'(e)]^2/s_{d^*}''(e)$.

To compare various PPAs, 
it is sufficient to look at $G_{d^*}(e_0)$:\cite{CV-PADE2} Indeed all Pade's have the same series around $e=0$, and if they have the same value at $e_0$, it is likely that their variations will be very similar.

\section{EVALUATING THE GROUNG-STATE ENERGY  $e_0$}
\label{APP-e0}
We now show how to evaluate the ground-state energy if unknown. We look for the value giving the highest number of similar PPAs. 
As mentioned in the previous appendix, it is sufficient to look at the values $G_{d^*}(e_0)$.
We define the quality of the result as
\begin{equation}
	Q_e(e_0)=\sum_{d_1^*}\sum_{d_2^*> d_1^*} \M\left(\frac{G_{d_1^*}(e_0)-G_{d_2^*}(e_0)}\epsilon\right),
\end{equation}
where $\M$ is a measure function as defined in Eq. (\ref{EQ-def-M}).
Unfortunately, this function may be discontinuous because the number of PPAs may eventually change. Then, the maximum of $Q_e(e_0)$ is found after a systematic search on a grid.

\newpage	
\begin{center}
{\Large \bf High-temperature series $J_1$-$J_2$-$J_d\,$-Heisenberg model on the kagome lattice.}
\end{center}

\section*{HT-series for $\X$}
The susceptibility HT-series polynomials are defined as:
\begin{eqnarray}
\label{EQ-defchip}
\label{EQ-defP}
	\frac{\X^{\HT}(T)T}{C}&=&\sum_{i=0}^n P_i(J_1,J_2,J_d) \beta^i\\
			&=&\sum_{i=0}^n \frac{ p_i(\nu_2,\nu_3)}{i!}\left(\frac{ J_1} {2T}\right)^i
\end{eqnarray}
where $\beta=1/T$, $\nu_2=J_2/J_1$ and $\nu_3=J_d/J_1$.
\begin{align}
	p_0(\nu_2,\nu_3)=&1\\
\label{EQ-p1}
	p_1(\nu_2,\nu_3)=&-2-2\,\nu_2-\nu_3\\
	p_2(\nu_2,\nu_3)=&4+8\,\nu_3+16\,\nu_2+8\,\nu_2\nu_3+4\,\nu_2^{2} \\
	p_3(\nu_2,\nu_3)=&-3\!-\!48\,\nu_3
		-81\,\nu_2-3\,\nu_2^{3}-48\,\nu_2^{2}\nu_3+2\,\nu_3^{3}
		-12\,\nu_2\nu_3^{2}
		-114\,\nu_2\nu_3-96\,\nu_2^{2}-12\,\nu_3^{2}\\
\nonumber
	p_4(\nu_2,\nu_3)=&-4+176\,\nu_3+172\,\nu_2-4\,\nu_2^{4}-32\,\nu_2\nu_3^{3}
		 +\!216\,\nu_2^{3}\nu_3+5\,\nu_3^{4}+260\,\nu_2^{2}\nu_3^{2}+396\,\nu_2\nu_3^{2}+1188\,\nu_2^{2}\nu_3
&\\&
		+\!432\,\nu_2^{3}\!-\!32\,\nu_3^{3}\!+\!1094\,\nu_2^{2}
					\!+\!1108\,\nu_2\nu_3\!+\!260\,\nu_3^{2}
\\
\nonumber
	p_5(\nu_2,\nu_3)=&-202+225\,\nu_3+605\,\nu_2-680\,\nu_2^{4}\nu_3-202\,\nu_2^{5}
				 -21\,\nu_3^{5}-270\,\nu_2^{2}\nu_3^{3}+60\,\nu_2\nu_3^{4}-3245\,\nu_2^{3}\nu_3^{2}
				-11430\,\nu_2^{3}\nu_3
&\\\nonumber&
				+60\,\nu_3^{4}
					-7970\,\nu_2^{2}\nu_3^{2}-410\,\nu_2\nu_3^{3}
				-1360\,\nu_2^{4}-270\,\nu_3^{3}
				-8465\,\nu_2\nu_3^{2}-17230\,\nu_2^{2}\nu_3
				-11645\,\nu_2^{3}-2935\,\nu_3^{2}
&\\&
				-7630\,\nu_2\nu_3-5595\,\nu_2^{2} &
\\
\nonumber
	p_6(\nu_2,\nu_3)=&1513+4104\,\nu_2-4206\,\nu_3-14253\,\nu_2^{2}
		+122658\,\nu_2\nu_3^{2}+187998\,\nu_2^{2}\nu_3+15912\,\nu_2\nu_3
		+132562\,\nu_2^{3}
&\\\nonumber&
		+13170\,\nu_3^{2}+20210\,\nu_3^{3}
		+98088\,\nu_2^{4}+251238\,\nu_2^{3}\nu_3
		+40632\,\nu_2\nu_3^{3}
		+183249\,\nu_2^{2}\nu_3^{2}-2646\,\nu_2\nu_3^{4}+104100\,\nu_2^{4}\nu_3
&\\\nonumber&
		+132630\,\nu_2^{3}\nu_3^{2}+38550\,\nu_2^{2}\nu_3^{3}-4170\,\nu_2^{2}\nu_3^{4}
		+18080\,\nu_2^{3}\nu_3^{3}
		+26370\,\nu_2^{4}\nu_3^{2}+3384\,\nu_2^{5}\nu_3
		+1104\,\nu_2\nu_3^{5}-4170\,\nu_3^{4}
&\\&
		+1104\,\nu_3^{5}+6768\,\nu_2^{5}
		+1513\,\nu_2^{6}
		-{\frac {399}{2}}\,\nu_3^{6}
\\
\nonumber
	p_7(\nu_2,\nu_3)=&13844-151620\,\nu_2-74704\,\nu_3+139083\,\nu_2^{2}
			-{\frac {1966153}{2}}\,\nu_2\nu_3^{2}-1293383\,\nu_2^{2}\nu_3
			+299964\,\nu_2\nu_3
			-{\frac {189371}{2}}\,\nu_2^{3}
&\\\nonumber&
			+139755\,\nu_3^{2}
			-371574\,\nu_3^{3}-{\frac {4759391}{2}}\,\nu_2^{4}
			-3589292\,\nu_2^{3}\nu_3
			-1154895\,\nu_2\nu_3^{3}-{\frac {6770421}{2}}\,\nu_2^{2}\nu_3^{2}
			-{\frac {82789}{2}}\,\nu_2\nu_3^{4}
&\\\nonumber&
			-3827614\,\nu_2^{4}\nu_3-3512439\,\nu_2^{3}\nu_3^{2}
			-1749132\,\nu_2^{2}\nu_3^{3}-86618\,\nu_2^{2}\nu_3^{4}-990822\,\nu_2^{3}\nu_3^{3}
			-1991059\,\nu_2^{4}\nu_3^{2}
&\\\nonumber&
			-775061\,\nu_2^{5}\nu_3+7651\,nu_2\nu_3^{5}
			-354109\,\nu_2^{4}\nu_3^{3}+938\,\nu_2\nu_3^{6}+{\frac {67319}{2}}\,\nu_2^{3}\nu_3^{4}
			-53018\,\nu_2^{6}\nu_3
			-3738\,\nu_2^{2}\nu_3^{5}
&\\&
			-{\frac {262773}{2}}\,\nu_2^{5}\nu_3^{2}
			+13615\,\nu_3^{4}-3738\,\nu_3^{5}
			-{\frac {1275309}{2}}\,\nu_2^{5}
			-106036\,\nu_2^{6}+938\,\nu_3^{6}+13844\,\nu_2^{7}+160\,\nu_3^{7}
\\
\nonumber
	p_8(\nu_2,\nu_3)=&-186286-137536\,\nu_2+1145568\,\nu_3
			+6324260\,\nu_2^{2}-2088608\,\nu_2\nu_3^{2}-4756328\,\nu_2^{2}\nu_3
			+857352\,\nu_2\nu_3
&\\\nonumber&
			-10846320\,\nu_2^{3}-1664724\,\nu_3^{2}
			+1483992\,\nu_3^{3}+16142374\,\nu_2^{4}+41699268\,\nu_2^{3}\nu_3
			+19842784\,\nu_2\nu_3^{3}
&\\\nonumber&
			+40719726\,\nu_2^{2}\nu_3^{2}
			+6650764\,\nu_2\nu_3^{4}+67176864\,\nu_2^{4}\nu_3
			+82220052\,\nu_2^{3}\nu_3^{2}
			+42915800\,\nu_2^{2}\nu_3^{3}
			+10425242\,\nu_2^{2}\nu_3^{4}
&\\\nonumber&
			+46137688\,\nu_2^{3}\nu_3^{3}
			+65417170\,\nu_2^{4}\nu_3^{2}+54159980\,\nu_2^{5}\nu_3
			-1232896\,\nu_2\nu_3^{5}
			+18554568\,\nu_2^{4}\nu_3^{3}
			+127572\,\nu_2\nu_3^{6}
&\\\nonumber&
			+6383884\,\nu_2^{3}\nu_3^{4}+3657208\,\nu_2^{6}\nu_3-1281328\,\nu_2^{2}\nu_3^{5}
			+26510312\,\nu_2^{5}\nu_3^{2}
			+2892272\,\nu_3^{4}-870832\,\nu_3^{5}
			+32058256\,\nu_2^{5}
&\\\nonumber&
			+4069692\,\nu_2^{6}+248420\,\nu_3^{6}
			+764352\,\nu_2^{7}-61008\,\nu_3^{7}-186286\,\nu_2^{8}
			+11421\,\nu_3^{8}+248420\,\nu_2^{2}\nu_3^{6}-592716\,\nu_2^{3}\nu_3^{5}
&\\&
			+1625812\,\nu_2^{4}\nu_3^{4}+4060012\,\nu_2^{5}\nu_3^{3}
			+811044\,\nu_2^{6}\nu_3^{2}
			-61008\,\nu_2\nu_3^{7}+382176\,\nu_2^{7}\nu_3
\end{align}
\begin{align}
\nonumber
	p_9(\nu_2,\nu_3)=&-2329677+26960814\,\nu_2+11526543\,\nu_3
		-56758545\,\nu_2^{2}+128462472\,\nu_2\nu_3^{2}
			+154668708\,\nu_2^{2}\nu_3
&\\\nonumber&
			-111054150\,\nu_2\nu_3
			-115342752\,\nu_2^{3}-36200385\,\nu_3^{2}
			+64998198\,\nu_3^{3}+319743243\,\nu_2^{4}
			-220814460\,\nu_2^{3}\nu_3
&\\\nonumber&
			-147995460\,\nu_2\nu_3^{3}
			-152834616\,\nu_2^{2}\nu_3^{2}-244445922\,\nu_2\nu_3^{4}
			-822646359\,\nu_2^{4}\nu_3-1365041772\,\nu_2^{3}\nu_3^{2}
&\\\nonumber&
			-706484844\,\nu_2^{2}\nu_3^{3}
			-436794300\,\nu_2^{2}\nu_3^{4}
			-1401175632\,\nu_2^{3}\nu_3^{3}
			-1793770353\,\nu_2^{4}\nu_3^{2}
			-1427828508\,\nu_2^{5}\nu_3
&\\\nonumber&
			+3083328\,\nu_2\nu_3^{5}-944463996\,\nu_2^{4}\nu_3^{3}+2878929\,\nu_2\nu_3^{6}
			-456663087\,\nu_2^{3}\nu_3^{4}
			-576786222\,\nu_2^{6}\nu_3
			+24409008\,\nu_2^{2}\nu_3^{5}
&\\\nonumber&
			-1226646711\,\nu_2^{5}\nu_3^{2}-81429093\,\nu_3^{4}+14618331\,\nu_3^{5}
			-580389138\,\nu_2^{5}
			-314815344\,\nu_2^{6}
			-3893763\,\nu_3^{6}
&\\\nonumber&
			-55709883\,\nu_2^{7}+1180962\,\nu_3^{7}
			+7338168\,\nu_2^{8}-311904\,\nu_3^{8}+7769322\,\nu_2^{2}\nu_3^{6}
			-13919346\,\nu_2^{3}\nu_3^{5}-172316772\,\nu_2^{4}\nu_3^{4}
&\\\nonumber&
			-380586006\,\nu_2^{5}\nu_3^{3}
			-287537220\,\nu_2^{6}\nu_3^{2}
			-965718\,\nu_2\nu_3^{7}-27272862\,\nu_2^{7}\nu_3
			+1180962\,\nu_2^{2}\nu_3^{7}-5141667\,\nu_2^{3}\nu_3^{6}
&\\\nonumber&
			+9044406\,\nu_2^{4}\nu_3^{5}
			-59040405\,\nu_2^{5}\nu_3^{4}
			-33586860\,\nu_2^{6}\nu_3^{3}-20688669\,\nu_2^{7}\nu_3^{2}
			-311904\,\nu_2\nu_3^{8}+3669084\,\nu_2^{8}\nu_3
&\\&
			-2329677\,\nu_2^{9}
			+37370\,\nu_3^{9}
\end{align}
\begin{align}
\nonumber
	p_{10}(\nu_2,0)=&44494564+32699900\,\nu_2-1607336300\,\nu_2^2
	+4682885400\,\nu_2^3-1969984450\,\nu_2^4
	-3722864284\,\nu_2^5
&\\&
	+12819641560\,\nu_2^6
	+2641862210\,\nu_2^7+853326455\,\nu_2^8
	-109501560\,\nu_2^9+44494564\,\nu_2^{10}
\\
\nonumber
	p_{10}(0,\nu_3)=&44494564-323940580\,\nu_3+753910650\,\nu_3^2
	-746061580\,\nu_3^3+248953155\,\nu_3^4
				+673343648\,\nu_3^5
&\\&
			-251041900\,\nu_3^6
				+86401210\,\nu_3^7-21228160\,\nu_3^8
				+4486600\,\nu_3^9
				-\frac{1698455}2\,\nu_3^{10}
\\
\nonumber
	p_{11}(0,\nu_3)=&568071766-\frac{5959595279}2\,\nu_3
	+\frac{19215984161}2\,\nu_3^2
	-\frac{49178193933}2\,\nu_3^3
				+33878328495\,\nu_3^4-30148742943\,\nu_3^5
&\\&
				+8479066530\,\nu_3^6-2914976526\,\nu_3^7
				+\frac{1749626615}2\,\nu_3^8
				-246955709\,\nu_3^9
				+58891206\,\nu_3^{10}-8569254\,\nu_3^{11}\\
	p_{12}(0,0)=&-15809083611\\
	p_{13}(0,0)=&-\frac{386791997479}2\\
	p_{14}(0,0)=&7857174705265\\
	p_{15}(0,0)=&84970643937857\\
	p_{16}(0,0)=&-5176017551551181\\
\end{align}
\newpage
\section*{HT series for $c_V^\spin$}
The specific heat HT-series polynomials are defined as:
\begin{eqnarray}
\label{EQ-CVHTdef}
\label{EQ-defQ}
	c_V^\spin(T) &=&\sum_{i=2}^n Q_i(J_1,J_2,J_d)\beta^i\\
		&=& \sum_{i=2}^n \frac{q_i(\nu_2,\nu_3)}{i!} \left(\frac{J_1}{2T}\right)^i
\end{eqnarray}
\begin{align}
q_0(\nu_3,\nu_3)=&0\\
q_1(\nu_3,\nu_3)=&0\\
\label{EQ-q2}
q_2(\nu_3,\nu_3)=&3+3\,\nu_2^2+\frac{3}{2}\,\nu_3^2\\
q_3(\nu_3,\nu_3)=&-27\,\nu_2-54\,\nu_3\,\nu_2+\frac{9}{2}\,\nu_3^3\\
q_4(\nu_3,\nu_3)=&-153+144\,\nu_2+108\,\nu_3+144\,\nu_3\,\nu_2
	-90\,\nu_3^2
	+252\,\nu_3\,\nu_2^2-72\,\nu_3^2\,\nu_2-153\,\nu_2^4
	-90\,\nu_3^2\,\nu_2^2-\frac{45}{2}\,\nu_3^4&\\
\end{align}
\begin{align}
\nonumber
q_5(\nu_3,\nu_3)=&3300\,\nu_2-900\,\nu_3-2850\,\nu_2^2+5700\,\nu_3\,\nu_2
	-150\,\nu_3^2+1650\,\nu_2^3-4650\,\nu_3\,\nu_2^2+750\,\nu_3^2\,\nu_2
	-450\,\nu_3^3
&\\&
	+6000\,\nu_3\,\nu_2^3-1350\,\nu_3^2\,\nu_2^2
	+3300\,\nu_3^3\,\nu_2-300\,\nu_3^2\,\nu_2^3
		-450\,\nu_3^3\,\nu_2^2-225\,\nu_3^5
\\
\end{align}
\begin{align}
\nonumber
q_6(\nu_3,\nu_3)=&32085/2-24570\,\nu_2-20655\,\nu_3-21735\,\nu_2^2
	+1620\,\nu_3\,\nu_2+\frac{33615}{2}\,\nu_3^2+19890\,\nu_2^3
	-131085\,\nu_3\,\nu_2^2+43200\,\nu_3^2\,\nu_2
&\\\nonumber&
	-9225\,\nu_3^3
	+9450\,\nu_2^4+46980\,\nu_3\,\nu_2^3
	-81675\,\nu_3^2\,\nu_2^2
	-2430\,\nu_3^3\,\nu_2+\frac{8235}{2}\,\nu_3^4-32805\,\nu_3\,\nu_2^4+54000\,\nu_3^2\,\nu_2^3
&\\&
	-28080\,\nu_3^3\,\nu_2^2+19710\,\nu_3^4\,\nu_2
	+\frac{32085}{2}\,\nu_2^6
	+16875\,\nu_3^2\,\nu_2^4+4950\,\nu_3^3\,\nu_2^3
	+\frac{8235}{2}\,\nu_3^4\,\nu_2^2+\frac{945}{4}\,\nu_3^6
\\
\end{align}
\begin{align}
\nonumber
q_7(\nu_3,\nu_3)=&-10143-644301\,\nu_2+269892\,\nu_3
	+1052226\,\nu_2^2
	-934920\,\nu_3\,\nu_2-75411\,\nu_3^2
	-460404\,\nu_2^3+1893654\,\nu_3\,\nu_2^2
&\\\nonumber&
	-31752\,\nu_3^2\,\nu_2
	+122598\,\nu_3^3+119511\,\nu_2^4-1000629\,\nu_3\,\nu_2^3
	+242991\,\nu_3^2\,\nu_2^2-1032381\,\nu_3^3\,\nu_2-31311\,\nu_3^4
&\\\nonumber&
	-\frac{930069}{2}\,\nu_2^5
	+269010\,\nu_3\,\nu_2^4+827757\,\nu_3^2\,\nu_2^3
	-257544\,\nu_3^3\,\nu_2^2
	-\frac{125685}{2}\,\nu_3^4\,\nu_2+58653\,\nu_3^5
	-1421343\,\nu_3\,\nu_2^5
&\\\nonumber&
	-187425\,\nu_3^2\,\nu_2^4
	-980343\,\nu_3^3\,\nu_2^3+\frac{52479}{2}\,\nu_3^4\,\nu_2^2
	-227997\,\nu_3^5\,\nu_2
	-10143\,\nu_2^7+7938\,\nu_3^2\,\nu_2^5-38808\,\nu_3^3\,\nu_2^4
&\\&
	+66150\,\nu_3^4\,\nu_2^3+58653\,\nu_3^5\,\nu_2^2+\frac{57771}{4}\,\nu_3^7&
\\
\end{align}
\begin{align}
\nonumber
q_8(\nu_3,\nu_3)=&-2859213+6178704\,\nu_2+5449416\,\nu_3
	+9429168\,\nu_2^2-9235296\,\nu_3\,\nu_2-4847052\,\nu_3^2-20274912\,\nu_2^3
&\\\nonumber&
	+53705904\,\nu_3\,\nu_2^2
	-21687792\,\nu_3^2\,\nu_2
	+5012952\,\nu_3^3+9225132\,\nu_2^4
	-62253744\,\nu_3\,\nu_2^3
	+48280344\,\nu_3^2\,\nu_2^2
&\\\nonumber&
	+3022992\,\nu_3^3\,\nu_2-1229970\,\nu_3^4
	-1534176\,\nu_2^5
	+56331912\,\nu_3\,\nu_2^4-65766624\,\nu_3^2\,\nu_2^3+40230624\,\nu_3^3\,\nu_2^2
&\\\nonumber&
	-12512304\,\nu_3^4\,\nu_2
	+1025136\,\nu_3^5
	-626304\,\nu_2^6-7418544\,\nu_3\,\nu_2^5
	+43961232\,\nu_3^2\,\nu_2^4
	-8318688\,\nu_3^3\,\nu_2^3
&\\\nonumber&
	+23324616\,\nu_3^4\,\nu_2^2-130368\,\nu_3^5\,\nu_2
	-125076\,\nu_3^6
	+10753344\,\nu_3\,\nu_2^6-12882912\,\nu_3^2\,\nu_2^5+10512768\,\nu_3^3\,\nu_2^4
&\\\nonumber&
	-13091568\,\nu_3^4\,\nu_2^3
	+4212600\,\nu_3^5\,\nu_2^2
	-3452400\,\nu_3^6\,\nu_2
	-2859213\,\nu_2^8-3844932\,\nu_3^2\,\nu_2^6
	-741888\,\nu_3^3\,\nu_2^5
&\\&
	-2292990\,\nu_3^4\,\nu_2^4-654192\,\nu_3^5\,\nu_2^3
	-125076\,\nu_3^6\,\nu_2^2+\frac{89271}{2}\,\nu_3^8
\end{align}
\begin{align}
\nonumber
q_9(\nu_3,\nu_3)=&5600664+184435056\,\nu_2-100067400\,\nu_3
	-433650996\,\nu_2^2+227967048\,\nu_3\,\nu_2
	+78952644\,\nu_3^2
	+179433792\,\nu_2^3
&\\\nonumber&
	-607203540\,\nu_3\,\nu_2^2-42816600\,\nu_3^2\,\nu_2
	-63694188\,\nu_3^3+95703120\,\nu_2^4
	-144709416\,\nu_3\,\nu_2^3
	+485374032\,\nu_3^2\,\nu_2^2
&\\\nonumber&
	+276310440\,\nu_3^3\,\nu_2+35501328\,\nu_3^4
	+123924168\,\nu_2^5+441623340\,\nu_3\,\nu_2^4
	-1670214816\,\nu_3^2\,\nu_2^3
	+372083544\,\nu_3^3\,\nu_2^2
&\\\nonumber&
	-127644984\,\nu_3^4\,\nu_2-24001596\,\nu_3^5
	-70144056\,\nu_2^6+144516960\,\nu_3\,\nu_2^5
	+1235200104\,\nu_3^2\,\nu_2^4
	-631218096\,\nu_3^3\,\nu_2^3
&\\\nonumber&
	+54034452\,\nu_3^4\,\nu_2^2+115234488\,\nu_3^5\,\nu_2
	+10998828\,\nu_3^6+146195604\,\nu_2^7
	-127553616\,\nu_3\,\nu_2^6
	-684430884\,\nu_3^2\,\nu_2^5
&\\\nonumber&
	+852307920\,\nu_3^3\,\nu_2^4-593753004\,\nu_3^4\,\nu_2^3
	+250454268\,\nu_3^5\,\nu_2^2+4221396\,\nu_3^6\,\nu_2
	-8027748\,\nu_3^7
	+442063656\,\nu_3\,\nu_2^7
&\\\nonumber&
	+75600216\,\nu_3^2\,\nu_2^6+423090216\,\nu_3^3\,\nu_2^5
	+106291116\,\nu_3^4\,\nu_2^4+101241576\,\nu_3^5\,\nu_2^3
	+22410432\,\nu_3^6\,\nu_2^2
&\\\nonumber&
	+12159720\,\nu_3^7\,\nu_2
	+5600664\,\nu_2^9+4667544\,\nu_3^2\,\nu_2^7
	+32428512\,\nu_3^3\,\nu_2^6-21895272\,\nu_3^4\,\nu_2^5
	+2416392\,\nu_3^5\,\nu_2^4
&\\&
	-17600328\,\nu_3^6\,\nu_2^3
	-8027748\,\nu_3^7\,\nu_2^2-1161810\,\nu_3^9
\end{align}
\begin{align}
\nonumber
q_{10}(\nu2,0)=&\frac{559095695}2-2276071650\, \nu_2-4567367250\, \nu_2^2
	+14999658300\, \nu_2^3-11395200825\, \nu_2^4
	+3512454300\, \nu_2^5
&\\&
	-4851077175\, \nu_2^6
	+1271847150\, \nu_2^7
	-191251800\, \nu_2^8
	+\frac{1559095695}{2}\, \nu_2^{10}
\end{align}
\begin{align}
\nonumber
q_{10}(0,\nu_3)=&\frac{559095695}2-1968883875\, \nu_3+\frac{3707972775}{2}\nu_3^2
	-2766779100\, \nu_3^3+953949150\, \nu_3^4
	-929445975\, \nu_3^5
&\\&
	+73620225\, \nu_3^6
	-111223125\, \nu_3^7-\frac{59072625}{2}\, \nu_3^8
	-\frac{45148455}{4}\, \nu_3^{10}
\end{align}
\begin{align}
\nonumber
q_{11}(0,\nu_3)=&3188690010+49046431005\, \nu_3
	-\frac{127470667995}{2}\, \nu_3^2+\frac{103764854985}{2}\, \nu_3^3
	-41081190975\, \nu_3^4+16829360625\, \nu_3^5
&\\&
	-15771729270\, \nu_3^6
	+\frac{10254468675}{2}\, \nu_3^7
	-3039511530\, \nu_3^8+1205526630\, \nu_3^9
	+\frac{390259485}{4}\, \nu_3^{11}\\
	q_{12}(0,0)=&-\frac{603545755725}2.\\
	q_{13}(0,0)=&2163459683034\\
	q_{14}(0,0)=&\frac{627738503442687}4\\
	q_{15}(0,0)=&-\frac{3574008872020125}2\\
	q_{16}(0,0)=&-105051131047391805\\
	q_{17}(0,0)=&1788835670130700224
\end{align}
\newpage
\section*{Parameters at the highest quality fit for $\X$ at fixed $J_1$}
\begin{table}[h]
\begin{center}
\begin{tabular}{crr|rr|cc|c}
$J_1$ & \multicolumn{1}{c}{$J_2$} & \multicolumn{1}{c|}{$J_d$} & \multicolumn{1}{c}{$\theta$} & $J_{C_V}$ & $(A-1)$ & $B (K^{-1})$ &  \multicolumn{1}{c}{$Q_\X$} \\
     \multicolumn{1}{c}{$(K)$}      &  \multicolumn{1}{c}{$(K)$}       &   \multicolumn{1}{c|}{$(K)$}      &   \multicolumn{1}{c}{$(K)$}       &   \multicolumn{1}{c|}{$(K)$}          & $\times10^2\;$ & $\times 10^4$ &  \\
\hline
-19.5 & -9.438 & 0 &28.9&13.3&  1.0 & -0.60 & 7 \\
-14.2 & 0 & 12.212 &8.1&10.2&  2.7 & -1.00 & 10 \\
\hline
-24 & 13.951 & -6.488 &13.3&17.2&  -1.5 & 0.02 & 3.4 \\
-22 & 11.348 & -2.189 &11.8&15.2&  -0.4 & -0.24 & 6 \\
-20 & 8.868 & 1.538 &10.4&13.4&  0.8 & -0.55 & 6 \\
-20 & 12.034 & -5.256 &10.6&14.5&  1.1 & -0.68 & 5.8 \\
-18 & 6.594 & 4.459 &9.2&11.9&  1.9 & -0.82 & 7 \\
-16 & 2.155 & 10.404 &8.6&10.9&  2.3 & -0.94 & 7 \\
-14 & -1.266 & 13.729 &8.4&10.5&  2.5 & -0.96 & 7 \\
-12 & -4.203 & 15.820 &8.3&10.4&  2.5 & -0.96 & 8 \\
-10 & -6.744 & 16.972 &8.3&10.4&  2.6 & -0.99 & 8 \\
-8 & -8.933 & 17.274 &8.3&10.5&  2.5 & -0.97 & 8 \\
-6 & -10.450 & 16.643 &8.1&10.3&  2.7 & -1.02 & 8 \\
-4 & -11.769 & 15.484 &8.0&10.1&  2.8 & -1.04 & 8 \\
-2 & -12.825 & 13.793 &7.9&9.9&  2.9 & -1.05 & 9 \\
-0.2 & -13.236 & 11.602 &7.6&9.5&  3.1 & -1.12 & 9.8 \\
2 & -13.723 & 8.510 &7.5&9.3&  3.3 & -1.16 & 9.6 \\
4 & -14.082 & 5.204 &7.5&9.2&  3.3 & -1.17 & 7.9 \\
6 & -14.144 & 1.137 &7.6&9.4&  3.3 & -1.16 & 7 \\
8 & -14.239 & -3.725 &8.1&10.1&  2.7 & -1.00 & 9 \\
10 & -6.105 & -29.960 &11.1&14.8&  0.4 & -0.47 & 5 \\
12 & -13.207 & -18.525 &10.5&13.6&  0.7 & -0.49 & 6 \\
10 & -13.106 & -10.897 &8.6&11.1&  2.4 & -0.94 & 7 \\
8 & -3.242 & -31.180 &10.8&14.5&  0.6 & -0.52 & 6 \\
6 & -0.941 & -30.507 &10.2&13.7&  1.1 & -0.64 & 6 \\
4 & 1.323 & -31.075 &10.2&13.7&  1.1 & -0.63 & 8 \\
2 & 3.347 & -31.236 &10.3&13.7&  1.0 & -0.62 & 8 \\
-0.2 & 5.355 & -30.709 &10.2&13.7&  1.1 & -0.64 & 8 \\
-2 & 7.002 & -31.236 &10.6&14.2&  0.8 & -0.57 & 7 \\
-4 & 8.573 & -30.730 &10.8&14.5&  0.6 & -0.53 & 6 \\
-6 & 10.074 & -30.317 &11.1&15.0&  0.4 & -0.48 & 6 \\
-8 & 11.455 & -29.648 &11.4&15.4&  0.3 & -0.44 & 6 \\
-10 & 12.643 & -28.241 &11.5&15.7&  0.2 & -0.44 & 6 \\
-12 & 13.809 & -27.377 &11.9&16.3&  0.0 & -0.39 & 5 \\
-14 & 14.651 & -25.031 &11.9&16.5&  0.1 & -0.44 & 4.8 \\
-16 & 15.212 & -22.207 &11.9&16.6&  0.1 & -0.44 & 4.3 \\
-18 & 15.314 & -18.172 &11.8&16.5&  0.2 & -0.48 & 3.9\\
\hline
\hline
-12 & \multicolumn{1}{c}{15.6} & \multicolumn{1}{c}{4}& 8.2 & 10.3 & 2.7& -1.00 & 14.8
\end{tabular}
\end{center}
\caption{The Curie-Weiss temperature is $\theta=P_1(J_1,J_2,J_d)=-J_1-J_2-J_d/2$ (Eqs.\ref{EQ-defP},\ref{EQ-p1}) and the leading term of $C_V$ at high temperature $J_{C_V}^2=Q_2(J_1,J_2,J_d)=3/8 (J_1^2+J_2^2+J_d^2/2)]$ (Eqs.\ref{EQ-defQ},\ref{EQ-q2}).
$A$ and $B$ are defined in Eq.3 of the article.
The last line is at the best point for both $\X$ and $C_V$ and the last column means $Q_\X+Q_{C_V}^{}$.
\label{TAB}
}
\end{table}%

\end{document}